\documentclass[useAMS,usenatbib]{mn2e}

\usepackage{graphicx}
\usepackage{float}
\usepackage{upgreek}
\usepackage{amsmath}
\usepackage{breqn}
\usepackage{color,soul}
\usepackage{pdflscape}
\usepackage{afterpage}
\usepackage{capt-of}
\usepackage{stfloats}
\usepackage{caption}
\usepackage{natbib}
\usepackage{subfig}

\makeatletter

\newcommand{\Rmnum}[1]{\expandafter\@slowromancap\romannumeral #1@}
\makeatother
\def\apj{The Astrophysical Journal}

\title[The mass fraction of AGN and the Fundamental Plane of black hole activity]{The mass fraction of AGN and the Fundamental Plane of black hole activity from a large X-ray selected sample of LINERs}

\author[D. M. Nisbet \& P. N. Best]{D. M. Nisbet$^{1}$\thanks{E-mail: dmn@roe.ac.uk}, and P. N. Best$^{1}$\\
$^{1}$SUPA, Institute for Astronomy, Royal Observatory Edinburgh, Blackford Hill, Edinburgh, EH9 3HJ, UK}

\begin{document}

\date{Accepted 2015 October 19.  Received 2015 October 14; in original form 2015 July 14 }

\pagerange{\pageref{firstpage}--\pageref{lastpage}} \pubyear{2015}

\maketitle

\label{firstpage}

\begin{abstract}
A sample of 576 X-ray selected LINERs was constructed by combining data from the 3XMM-DR4 and SDSS-DR7 catalogues. The sample was used to investigate the fraction of galaxies hosting a LINER, finding that the fraction is a strong function of both stellar mass and black hole mass (increasing as $\mathrm{f_{LINER} \propto M^{1.6 \pm 0.2}_{*}}$ and $\mathrm{f_{LINER} \propto M^{0.6 \pm 0.1}_{BH}}$ respectively) and that it rises close to unity at the highest black hole masses and lowest X-ray luminosities. After obtaining radio flux densities from the FIRST survey, the sample was also used to investigate the Fundamental Plane of black hole activity -- a scale-invariant relationship between black hole mass, X-ray luminosity and radio luminosity that is believed to hold across at least nine orders of magnitude of mass. There are key advantages in using only LINERs for the derivation as these are the counterparts of the ``low-hard'' X-ray binaries for which the relationship is tightest. The Fundamental Plane was found to be $\rm{log\, \left(\frac{L_R}{erg/s}\right) } =  0.65^{+0.07}_{-0.07}\,\,  \rm{log}\left(\frac{L_X}{10^{42}erg/s}\right) + 0.69^{+0.10}_{-0.10}\,\,\rm{log}\left(\frac{M_{BH}}{10^8\,M_{\odot}}\right) + 38.35^{+0.10}_{-0.10}$. The scatter around the plane was $0.73 \pm 0.03$ dex, too large to suggest that the Fundamental Plane can be used as a tool to estimate black hole mass from the observables of X-ray and radio luminosity. The black hole mass scaling is sensitive to the slope of the mass -- velocity dispersion relation and, in order to achieve consistency with X-ray binaries, the analysis favours a steep gradient for this relationship, as found in recent research.
\end{abstract}


\section{Introduction}
It has long been recognised that some form of feedback needs to play a part in the process of galaxy evolution. \cite{whit78} were unable to reproduce the shape of the galaxy luminosity function when running simulations of hierarchical cosmological models and argued that there must be some mechanism that lowers the efficiency of galaxy formation. Attention initially focussed mainly on feedback processes that might explain the faint end of the luminosity function, and, in particular, the role of supernovae. Following a study by \cite{bens03} that examined the impact of different feedback mechanisms on the shape of the bright end of the galaxy luminosity function, feedback from active galactic nuclei (AGN) has received increasing attention.

Active galaxies have traditionally been placed into a number of categories, according to their luminosity, spectral properties and ratio of nuclear to host galaxy stellar light; the categories include Seyfert galaxies, quasars, QSOs, LINERs, Blazars and radio galaxies.  Progressively, however, the view has developed that AGN can be more usefully divided into just two fundamental types \citep[see the review by][]{heck14}, with the division dependent on how much material is being accreted onto the supermassive black hole (SMBH). The threshold value that marks the division is approximately 1\% of the Eddington limit; the exact fraction depends on, inter alia, the rate of spin of the SMBH and, especially, the viscosity of the accretion disk \citep{qiao}. 

Radiative-mode AGN accrete at more than the 1\% threshold. They are generally associated with lower-mass black holes residing in younger, growing galaxies. Infall onto the SMBH occurs through a geometrically-thin, but optically-thick, accretion disk that reaches in to within a few Schwarzschild radii of the SMBH \citep[for example,][]{shak}. A galaxy hosting a radiative-mode AGN is likely to contain a huge mass of cold molecular gas, implying that there is substantial star formation and high rates of accretion onto the SMBH in a radiatively efficient process (see discussion in Heckman \& Best 2014). Stellar feedback is important in such galaxies; by contrast, AGN feedback, arising from radiation-induced expulsion of gas and dust from the galaxy, is likely to be a significant factor only when the AGN becomes particularly powerful. At that time, the winds may be sufficiently powerful to blow away surrounding gas and dust, leading to the cessation of each of star formation, growth in the galaxy and growth in the AGN \citep[for example,][]{silk,fabi99,king}. It is believed that this may be the phase that links the mass of the central black hole with that of its host galaxy. Establishing a causal link between radiative-mode feedback and the quenching of star formation is, however, fraught with difficulties caused by typically high redshifts, obscuration, the brightness of an AGN relative to its host galaxy (the opposite problem to obscuration) and the need to disentangle the effect of AGN feedback from that caused by stellar activity. 

Jet-mode AGN tend to be linked with more massive black holes that are found in older galaxies, those whose morphologies consist of classical bulges and massive ellipticals. Gas in these galaxies tends to be hot and little star formation is occurring. The AGN is accreting at significantly below -- less than about 1\% of -- the Eddington limit in a radiatively inefficient process. The structure of a jet-mode AGN is believed to differ from that of a radiative-mode AGN (Heckman \& Best 2014, and references therein). The accretion disk is either absent or truncated in the inner regions; instead, there is a geometrically-thick structure, which leads to an inflow time that is much shorter than the radiative cooling time \citep[see the discussion in][]{ho}.

A characteristic property of radiatively-inefficient accretion flows is that they generate two-sided jets. The jets can extend for mega-parsecs, far outside the galaxy itself. It is believed that much of the energetic output of the SMBH is channelled into the jets and that their power is, in turn, eventually transferred into the surrounding inter-galactic medium (IGM) where it offsets the radiative cooling of the gas. This means that the energy of a jet-mode AGN is deposited into the same gas that fuels the AGN -- the ingredients are in place for a feedback cycle. 

The mechanism that produces the jets is believed to be related to magnetic fields and angular momentum, but is not yet fully understood. Given that black holes are characterised simply by their mass and spin, most of the ongoing processes should be scale-invariant \citep{hein}. The facts that jets from black holes which span nine orders of magnitude in mass are morphologically similar and that their core emission tends to display the same flat power-law spectrum provide supporting evidence. The scale invariance leads to the concept of the Fundamental Plane of black hole activity -- an empirical relationship between the black hole mass, radio luminosity (a probe for the radio jet) and X-ray luminosity. Black hole binaries in our own galaxy and AGN both lie upon the same relationship. The tightest relationship for X-ray binaries is seen for those in the ``low-hard'' state; this implies binaries with low Eddington fractions and so the appropriate counterparts with which comparisons should be made are jet-mode AGN such as LINERs. More detailed descriptions of the Fundamental Plane are provided by, for example, \citet{merl03} and \citet{saik}.

A leading hypothesis to explain the launching of the jets is that the mass and spin of a SMBH distort space and time, twisting magnetic field lines into a coil that propels material outwisards. There is observational support for the theory. Firstly, \citet{mars} studied a comparatively nearby -- at a distance of 300 million parsecs -- blazar, BL Lacertae, during a period of extreme outburst and observed material spiralling outwards, in line with predictions. Secondly, it follows that the greater the mass of a black hole, the greater the probability of jets (and, to a lesser extent, the greater the power of the jets). This has indeed been observed, with Best et al (2005) finding that radio-AGN activity is a very strong function of black hole mass. 

There is plenty of observational evidence to support the feedback theory for jet-mode AGN: for example, the gas surrounding a galaxy is hot, highly ionised and mostly transparent and jet-inflated cavities can be observed at radio and X-ray frequencies. Despite that, the details of the feedback mechanism remain unclear.

In summary, there appears broad agreement about the general picture of AGN feedback, but the details remain poorly understood. An important metric in the study of AGN feedback is the fraction of galaxies in the local and near-local universe with AGN activity. This fraction provides a basic measurement of the AGN duty cycle and so places a constraint on theories and simulations that model the impact of accretion onto a black hole on galaxy evolution. 

This research focuses on jet-mode AGN. A dataset of 576 LINERs was constructed using the SDSS, 3XMM and FIRST catalogues, with the AGN selected on the basis of their hard (2 -- 12 keV) X-ray luminosity. X-rays provide one of the most direct evidences of nuclear activity and are fundamental to the study of the accretion process. This sample of X-ray selected LINERs was used to investigate three of the issues discussed above. 

The paper is structured as follows. Section 2 gives a brief description of the catalogues used to construct the sample of X-ray selected LINERs. Section 3 describes how the sample was constructed. Section 4 sets out the methodology used to calculate the mass-dependent fraction of galaxies hosting an X-ray LINER and shows the results. Section 5 derives the Fundamental Plane relationship for the sample of LINERs and then tests the Fundamental Plane derived against a sample of X-ray binaries. In Section 6, the Fundamental Plane was used to convert from an X-ray-selected to a pseudo radio-selected sample of LINERs, the mass-dependent fraction of galaxy hosts was re-examined, and then this was compared with determinations from radio-selected galaxies. Section 7 discusses the implication of the results obtained in the analysis. Section 8 summarises the key conclusions.

The cosmogeny used in this paper assumes parameter values of $\Omega_{m}$ = 0.3, $\Omega_{\Lambda}$ = 0.7 and $\rm{H_{0} = 70\, km\,s^{-1}\,Mpc^{-1}}$.

\section{The Catalogues}
This research has been carried out on a database of LINERs, for which both X-ray and radio luminosity information and black hole mass estimates are available. The sample was constructed by crossmatching the fourth data release of the 3XMM Serendipitous Source catalogue of X-ray sources (released by the XMM-Newton Survey Science Centre; XMM-SSC 2013) with the seventh data release from the Sloan Digital Sky Survey \citep[SDSS-DR7;][]{sdss} and then adding in radio luminosity data obtained from the Faint Images of the Radio Sky at Twenty-cm (FIRST) catalogue \citep{first}.

\subsection{The SDSS Spectroscopic Sample}  
The SDSS \citep{york} is a major multi-filter optical imaging and spectroscopic survey conducted on a dedicated 2.5-metre telescope at the Apache Point Observatory, New Mexico. This research used the seventh data release (SDSS-DR7), produced from a survey area that covers approximately one-quarter of the extragalactic sky. The spectra span an observed wavelength range of 3800 -- 9200 $\mathrm{\mathring{A}}$. 

The research used only galaxies selected to be in the main galaxy sample. The selection process of these is described in \cite{stra}. In summary, the objects must have a signal / noise of more than 5, they must not exceed certain saturation, brightness or blending thresholds, they must have a Petrosian magnitude of $r < 17.77$ and they must be spectroscopically confirmed as galaxies. The median redshift of galaxies in the main sample is 0.104.

For each of the galaxies in the catalogue, a large number of physical parameters have been estimated from the photometric and spectroscopic information in the MPA-JHU value-added catalogues \citep{brin}. The parameters used within this research are the redshift, total stellar mass, 4000$\mathrm{\mathring{A}}$ break strength, H$\updelta$ absorption measure, accurate emission-line fluxes and galaxy velocity dispersion from which the black hole mass is estimated (see Section 3.2). The methods used to derive the parameters are described in \citet{kauf03a,kauf03b,kauf03c} and \citet{heck04}.

\begin{center}
\begin{table*}
    \begin{tabular}{| l | l | r | r |}
    \hline
3XMM-DR4 & &  Sources Remaining &\\
Total Number of X-ray Sources & &  531261 &\\
After removing: & Non-unique Sources &   372728 &\\
& Extended Sources &   332637 &\\
& Unreliable Detections &   290396 &\\
& Sources with insufficient signal in either band &   258675 &\\ \hline
SDSS-DR7 & & &\\
Total Number of Galaxies & &  914694 &\\
After removing: & Galaxies not in Main Sample &   725094 &\\
& Galaxies with Redshift z $>$ 0.30 &   723916 &\\ \hline
Cross-match 3XMM-DR4 with SDSS-DR7 & &  1172 &\\
After removing: & Galaxies with no Black Hole Mass Data &   1094 &\\ \hline\hline
Split into Categories & & Classified & Selected in\\
& & via BPT & LINER sample\\
& LINERs &  170 & 170\\
& Unclassified Objects & 363 & 340\\
& Transition Objects & 157 & 45\\
& Narrow-line AGN & 40 & 34\\
& Seyferts  & 214 & 0\\
& Star-forming Galaxies & 150 & 0 \\
& Total  & 1094 & 589\\ \hline
After removing: & LINERs with no Radio Flux Data & &  576\\ \hline

    \end{tabular}
    \caption{A summary of the filtration process used to construct the sample of LINERs (see text for details).}\label{MyTable}
   \end{table*}
   \end{center}

\subsection{The 3XMM Catalogue of X-ray Sources}
3XMM-DR4 is the latest catalogue of serendipitous X-ray sources from the European Space Agency's (ESA) XMM-Newton observatory, and has been created by the XMM-Newton Survey Science Centre (SSC) on behalf of ESA. The catalogue contains source detections drawn from a total of 7427 observations taken by the XMM-Newton European Photon Imaging Camera (EPIC) between 2000 and 2012 inclusive. For net exposure time  $\geq$  1ksec, the total area of the catalogue fields is $\sim$1397 square degrees. Taking account of the substantial overlaps between observations, the net sky area covered independently is $\sim$794 square degrees.

The catalogue contains 531,261 detections, of which 372,728 are unique X-ray sources. Due to intrinsic features of the instrumentation as well as some shortcomings in the source detection process, some detections are considered to be spurious or their parameters are considered to be unreliable. Just over 80\% of the sources are considered to be clean (marked in the catalogue with a summary flag $< 3$).

This research is interested in AGN that are selected in the hard energy range. In this energy range, LINERs are dominated by a hard power-law component that shows little intrinsic absorption, whereas at lower energies ($\leq$ 2keV) an additional soft component is often present \citep[for example,][and references therein]{ho}. The 3XMM catalogue provides fluxes in the 2.0 -- 4.5 keV and 4.5 -- 12.0 keV bands. The median flux obtained by summing these two bands is $\sim 1.3 \times 10^{-14}$ $\mathrm{erg/cm^{2}/s}$, and around 20\% of sources have fluxes below $1.0 \times 10^{-14}$ $\mathrm{erg/cm^{2}/s}$. The flux values from the three EPIC cameras are, overall, in agreement to $\sim$10\%. The positional accuracy of the catalogue point source detections is generally $< 3$ arcseconds (90\% confidence radius) and 90\% of point sources have 1-sigma positional uncertainties  $< 2.4$ arcseconds.

\subsection{Faint Images of the Radio Sky at Twenty-cm}
FIRST is a project designed to produce a catalogue of radio sources from a survey area of over 10,000 square degrees \citep{first}. Approximately 80\% of the sky observed is in the north Galactic cap, and 20\% is in the south Galactic cap. Both regions are covered by the Sloan Digital Sky Survey. Using the Very Large Array (VLA) in its B-configuration, an atlas of maps was produced by coadding the twelve images adjacent to each pointing centre. These maps have 1.8 arcseconds pixels, a typical rms of 0.16 mJy, and an angular resolution of 5 arcseconds. The noise in the coadded maps varies by only 15\% from the best to the worst places in the maps, except in the vicinity of bright sources ($> 100$ mJy) where sidelobes can lead to an increased noise level. At the 1 mJy source detection threshold, there are $\sim$90 sources per square degree, $\sim$35\% of which have resolved structure on scales from 2 -- 30 arcseconds. The astrometric reference frame of the maps is accurate to 0.05 arcseconds, and individual sources have 90\% confidence error circles of radius $< 0.5$ arcseconds at the 3 mJy level and 1 arcsecond at the survey threshold. 

\section{The Construction of the Sample of LINERs}
The construction of the database of LINERs is described below and is summarised in Table 1.

\subsection{Parent X-ray Sample}
Three filters were applied to the list of 372,728 unique sources in the 3XMM-DR4 catalogue.

First, extended X-ray sources, defined as those with an extent radius of more than 6 arcseconds, were excluded in order to remove sources dominated by the Bremsstrahlung emission of the hot gas in galaxy groups and clusters. 11\% of the sources are extended, leaving 332,637 point sources.

Second, those X-ray sources whose detections were considered to be unreliable (those with a summary flag of higher than two) were filtered out; this left 290,396 X-ray sources.

Third, this research was interested in selecting LINERs, in particular looking at the hard X-ray emission from the AGN. Accordingly, the selection criterion was based on hard X-ray flux, obtained by summing the fluxes in the 2.0 - 4.5 keV and 4.5 - 12.0 keV bands. Sources were accepted only if there was a reliable signal in both of these X-ray bands.  258,675 galaxies passed this test. Results were also tested using stricter signal / noise criteria and the results were found to be consistent within the errors.

In order to facilitate comparisons with literature values, the analysis defined the X-ray flux to be that in the rest-frame 2.0 -- 10.0 keV range. It was assumed that the X-ray spectrum followed a $\upnu^{- \upalpha}$ relationship with $\upalpha = 1$; hence the 2.0 -- 12.0 keV flux was reduced by 10\% to convert it onto the 2.0 -- 10.0 keV range. A redshift K-correction was also applied.   

\subsection{Matching with the SDSS Catalogue}
There are 914,694 galaxies in the MPA-JHU SDSS-DR7 catalogue. Two cuts were applied to this list. As mentioned above, it was decided to include only sources that are in the main galaxy sample in order to avoid the introduction of biases into the sample of LINERs; for example, the luminous red galaxy sample, which probes only a range of galaxy parameter space, was thus excluded. This restriction reduced the number of galaxies by 21\%, leaving 725,094. To ensure that evolutionary effects did not distort the analysis, galaxies at a redshift of over 0.30 were rejected. A further 1178 galaxies failed this test, leaving 723,916. The analysis was also carried out using narrower redshift ranges and no significant differences in results were found.

Three positional cross-matches of the 3XMM and SDSS catalogues were performed; these used separation distances of less than or equal to 3 arcseconds, 5 arcseconds or 8 arcseconds as the criterion for deciding whether a successful match was found. They generated 980, 1172 and 1338 cross-matches respectively. The probabilities of a false match, estimated by randomising the coordinates of the X-ray sources and then performing the cross match, were found to be 2.1\%, 4.1\% and 8.0\% respectively. Given the need to maximise the number of cross-matches but to minimise the number of false detections, a separation distance of less than or equal to 5 arcseconds was selected as the most suitable criterion for deciding whether a successful match was found. For comparison, the positional accuracy of the 3XMM catalogue is under 3 arcseconds as measured by the 90\% confidence radius. The research was checked using separation distances of both 3 arcseconds and 8 arcseconds in the cross-matching process in order to ensure that the results were not dependent on the choice of separation distance; no significant differences were found.

Black hole masses for the galaxies were determined from the velocity dispersions, using the traditional correlation in the form:

\begin{equation} 
\mathrm{log\left(\frac{M_{BH}}{M_{\odot}}\right)} = \upalpha + \upbeta \, \mathrm{log}\left(\frac{\upsigma}{ 200\, \mathrm{km / s}}\right)
\end{equation}
Estimates of the intercept, $\upalpha$, and the slope, $\upbeta$, have varied. \citet{trem02} found values of 8.13 $\pm$ 0.06 and 4.02 $\pm$ 0.32 respectively and these values have been used in much of the research conducted with the SDSS catalogue. More recent estimates have tended to find higher values of beta; for example, \citet{mcco13} estimated the intercept and slope to be 8.32 $\pm$ 0.05 and 5.64 $\pm$ 0.32 respectively. In this research we have adopted the McConnell \& Ma values throughout when calculating the black hole masses, and in one part of the analysis (see Section 5.2) we show that the results arising from the use of the McConnell \& Ma relationship provide a better match when extended to X-ray binaries than would be obtained by using the Tremaine et al relationship.

78 of the 3XMM-SDSS matches were removed because no reliable velocity dispersion, and hence no black hole mass, information was available. This left 1094 sources.

\subsection{Diagnostic Tests to Select LINERs}
LINERs are characterised by weak, small-scale radio jets, luminosities that are weaker than Seyfert galaxies and quasars, and optical spectra that are dominated by emission lines from low-ionisation species -- primarily [O\Rmnum{1}]  $\uplambda$630\rm{0, [O\Rmnum{2}] $\uplambda\uplambda$3726,9 and [S\Rmnum{2}] $\uplambda\uplambda$6717, 31 \citep{ho97}. A classification scheme using optical emission line ratios was first put forward by \citet{heck80} and subsequently amendments have been proposed by, inter alia, \citet{bald81}, \citet{kewl01,kewl06} and \citet{kauf03a}. In this research, a two-pronged approach was adopted to identify LINERs.

The sources were first categorised, where possible, following the emission line ratio diagnostic method described in detail by Kewley et al (2006; their Section 3). Three diagnostic tests based on pairs of emission line ratios were prepared; the tests compared log([O\Rmnum{3}]/H$\upbeta$) with (i) log([N\Rmnum{2}]/H$\upalpha$), (ii) log([S\Rmnum{2}]/H$\upalpha$) and (iii) log([O\Rmnum{1}]/H$\upalpha$).  The sources were split into six categories based on their locations in these diagrams (or by the absence of the relevant lines). In those cases where the three tests produced different classifications, the general principle used was to give the greatest precedence to the third test and the least precedence to the first test. The 1094 sources were found to divide into 150 star-forming galaxies, 214 Seyferts, 170 LINERs, 157 transition objects (sources with characteristics intermediate  between those of star-forming galaxies and those of LINERs and Seyferts), 40 narrow-line AGN (a mixture of type 2 Seyferts and LINERs, but with insufficient data to permit a classification) and 363 objects for which an unambiguous classification was not possible due to the absence of sufficient emission lines. Those objects classified as LINERs were accepted and those classified as Seyferts or star-forming galaxies were rejected. There are likely to be large numbers of LINERs within the transition objects, narrow-line AGN and unclassified objects and so objects in these three categories were subjected to two further diagnostic tests to sift out the LINERs. 

\begin{figure}
\centering

\subfloat{%
  \centering
  \includegraphics[width=1.10\linewidth, height = 160pt]{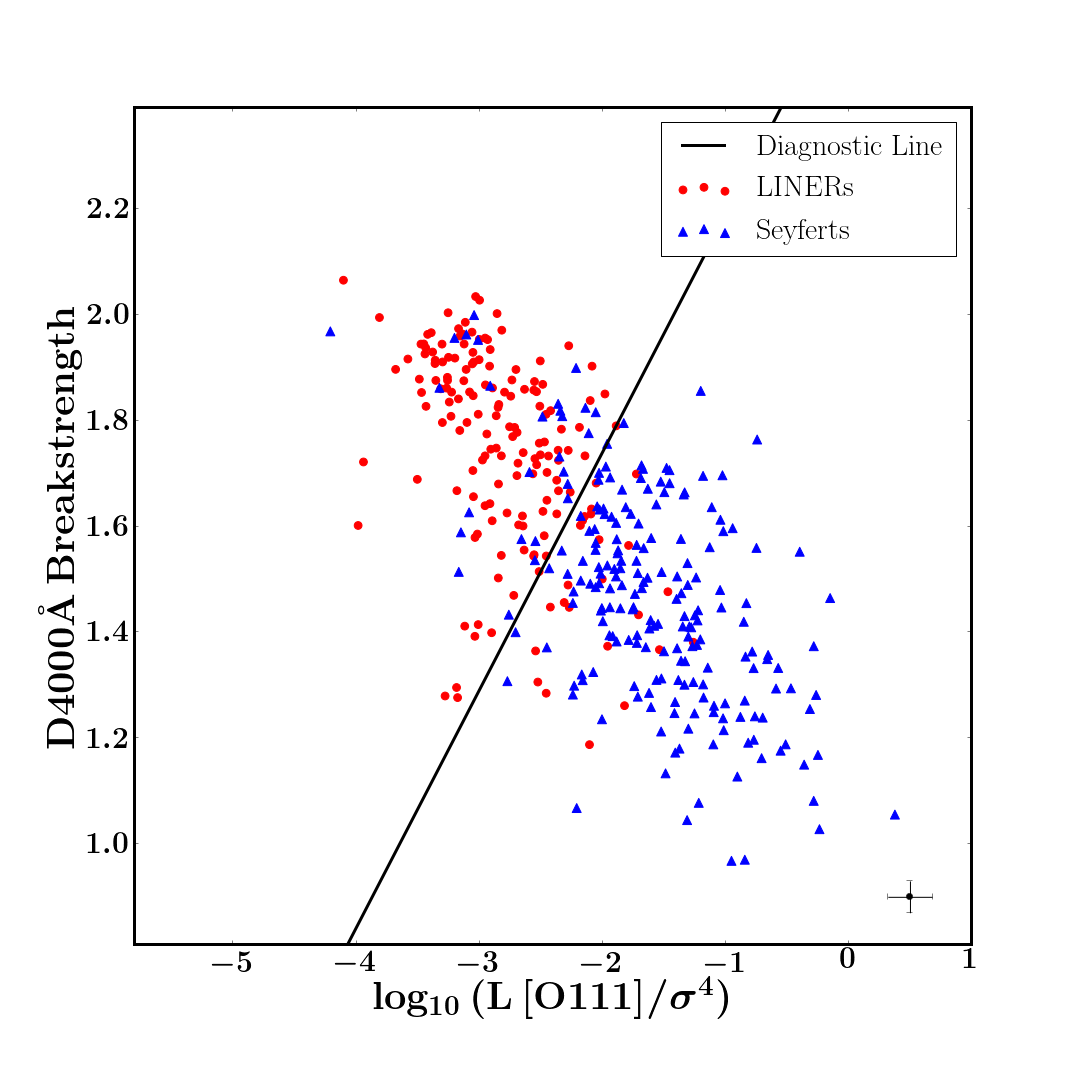} 
}

\subfloat{%
  \centering
  \includegraphics[width=1.10\linewidth, height = 160pt]{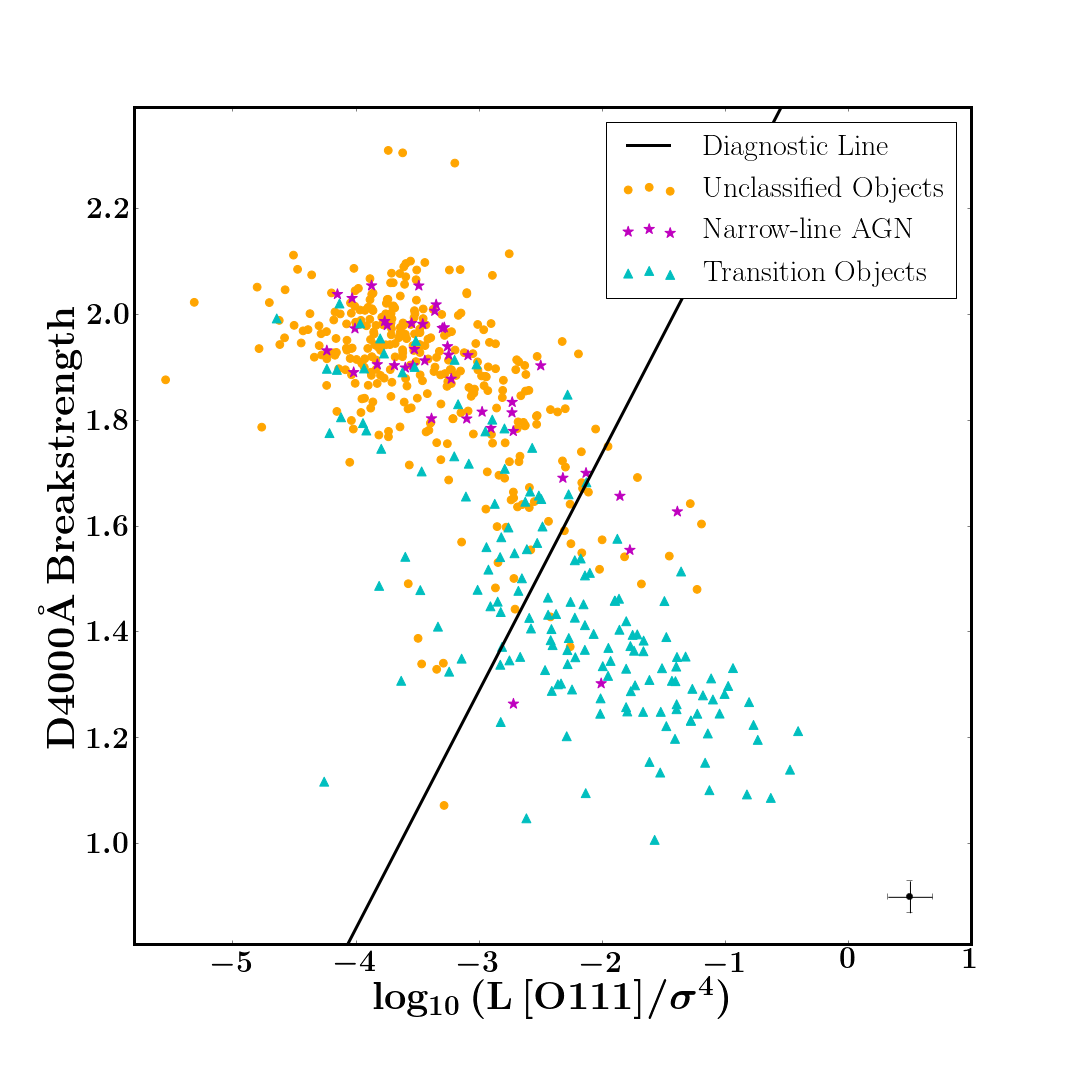} 
}

\caption{Using a diagnostic based on the 4000$\rm{\mathring{A}}$ breakstrength vs log(L[O\Rmnum{3}]/$\upsigma^{4}$) to identify LINERs amongst the unclassified objects, with L[O\Rmnum{3}]/$\upsigma^{4}$ measured in solar luminosities /$\rm{(km/s)^{4}}$. The separating line is defined from the top panel according to Equation 3 and is used to classify sources in the lower panel. Typical error bars are shown in the bottom right of the diagrams.} \label{fig:1}
\end{figure}

Firstly, the H$\upalpha$ luminosity was compared with the X-ray luminosity in order to separate out star-forming galaxies from the AGN within these three categories. This test exploits the fact that for star-forming galaxies the X-ray emission is expected to scale with H$\upalpha$ emission \citep[for example,][]{kenn94} and the ratio of X-ray emission to H$\upalpha$ emission is likely to be considerably lower than for LINERs. Sources were considered to be star-forming galaxies, and so rejected, if:
\begin{equation} 
\mathrm{log\, L}_{\mathrm{H}\alpha} >\mathrm{ 0.525\,log\,L_{X} - 12.6}
\end{equation}
The diagnostic line was set by deriving the upper limit to the star-forming population (the 150 star-forming galaxies) in our sample and, given that a second test would follow, was set at a conservative level. The test could be carried out only for those sources for which H$\alpha$ information was available. Out of these, 73 sources failed the test and were rejected: 6 unclassified objects, 66 transition objects and 1 narrow-line AGN. 

Secondly, the 4000$\mathrm{\mathring{A}}$ breakstrength was compared with log(L[O\Rmnum{3}]/$\upsigma^{4}$), where $\upsigma$ is the galaxy velocity dispersion; L[O\Rmnum{3}]/$\upsigma^{4}$ is a proxy for the ratio of the luminosity to the Eddington luminosity \citep{kewl06} and is therefore a fundamental property that divides LINERs from Seyferts since jet-mode and radiative-mode AGN separate in Eddington-scaled accretion rates (as discussed in the Introduction). The 4000$\mathrm{\mathring{A}}$ breakstrength provides a measure of specific star formation within a galaxy, which also typically differs between LINERs and Seyferts. A comparison of the breakstrength against the proxy for the Eddington ratio should, therefore, provide an effective means of identifying LINERs from the unclassified objects, transition objects and narrow-line AGN: the LINERs should be characterised by a relatively high breakstrength and low Eddington ratio. There is, indeed, a comparatively clean demarcation between those X-ray AGN classified as LINERs and Seyferts (Figure 1, upper). An optimisation procedure was used to select the most suitable diagnostic line -- that which maximises the completeness of the sample of LINERs and minimises contamination rates. Specifically, the division line was optimised to minimise $\rm{f_{wrong}}$, where $\rm{f^{2}_{wrong} = f^{2}_{wrong-LINER}+f^{2}_{wrong-Seyfert}}$, and $\rm{f_{wrong-LINER}}$ and $\rm{f_{wrong-Seyfert}}$ are the fractions of wrongly-diagnosed LINERs and Seyferts respectively. The diagnostic line was found to be:
\begin{equation} 
\mathrm{D_{4000}} =\mathrm{ 0.45\,log}\,\left(\frac{\mathrm{L[O\Rmnum{3}]}}{\sigma^{4}}\right)\,+ 2.64\,,
\end{equation}
where $\mathrm{\frac{L[O\Rmnum{3}]}{\upsigma^{4}}}$ is measured in Solar luminosities / $\rm{(km/s)^4}$.
Using the diagnostic line, it can be seen that the majority of unclassified objects can be categorised as LINERs as well as a majority of narrow-line AGN and minority of transition objects (Figure 1, lower). This is consistent with the facts that the narrow-line AGN tend to be in the same area as the LINERs in the BPT diagrams (although not unambiguously so) and the unclassified objects are mostly weak-lined and hence likely to be LINERs, whereas the transition objects have associated star formation and are more likely to be Seyferts.

After both diagnostic tests, the number of selected objects was found to be 589: all 170 LINERs, 340 out of 363 unclassified objects, 45 out of 157 transition objects and 34 out of 40 narrow-line AGN (see Table 1).

As discussed above, a cross-match with the SDSS catalogue was conducted using randomised coordinates for the X-ray sources and this generated 48 cross-matches; black hole mass data were available for 42 of these. These 42 sources were subjected to the same diagnostic tests and the results are shown in Table 2. 25 LINERs were identified, of which 24 had valid radio flux data. That represented 57$\%$ of the original 42 sources. For comparison, the corresponding proportion was 53$\%$ for the main LINER sample. The overall contamination of the LINER sample due to mismatches is $\leqslant 4\%$.

\begin{center}
\begin{table}
    \begin{tabular}{| l | r | r |}
    \hline
Categorised into: & via BPT &  of which LINERs \\
LINERs &  5 & 5\\
Unclassified Objects & 22 & 18\\
Transition Objects & 3 & 0\\
Narrow-line AGN & 2 & 2\\
Seyferts  & 0 & 0\\
Star-forming Galaxies & 10 & 0 \\
Total  & 42 & 25\\ \hline

    \end{tabular}
    \caption{The classification of a sample generated by a randomised cross-match.}\label{MyTable1}
   \end{table}
   \end{center}

\subsection{Radio Properties} 
Radio flux densities for the sources were then obtained from the FIRST Survey. Radio data were unavailable for 13 of the 589 galaxies, either because they were located outside of the FIRST coverage (there is not a perfect overlap with the SDSS survey area) or because the galaxies were located in regions of noisy radio flux close to bright sources. Catalogue values existed for 104 of the 576 LINERs. Radio flux densities for the other 472 LINERs were obtained by extracting image cut-outs from FIRST at the location of each source, fitting these with a Gaussian of FWHM 5 arcseconds at the location of the SDSS host galaxy and determining the best-fit normalisation. It was checked that this approach produced results that were consistent with the catalogue values, where they existed; it was also checked that the approach was consistent with using just the peak value. 

The radio flux density was converted to an integrated luminosity using the calculated redshift for each source, the assumed cosmogeny and the formula $\mathrm{L_{R}}$ = $\mathrm{\upnu}$ $\mathrm{L_{\upnu}}$, where $\upnu$ = 1.4 GHz for the FIRST data.

\subsection{Final Sample}
The final sample, therefore, consists of 576 LINERs. X-ray, optical and radio data, as well as estimates of redshift, stellar mass and black hole mass, are available for each of these objects. The properties of a subsample of 20 LINERs are set out in Table 4 in the Appendix. Data for all 576 LINERs are available electronically.

\section{Fraction of Galaxies that host LINERs}
The fraction of galaxies that host LINERs was investigated as a function of both stellar and black hole mass. 

\subsection{Areal Sky Coverage}
The 3XMM catalogue and SDSS-DR7 catalogue cover different areas of sky. The 3XMM pointings cover a net sky area of 794 square degrees. The SDSS-DR7 spectroscopy covers 8032 square degrees, but a small part of that area corresponds to areas of noisy or absent radio data. That proportion was estimated from the fact that radio data were unavailable for 13 out of 589 LINERs, suggesting that the ``missing'' area is approximately 2.2\%. The effective area was, therefore, taken to be 7855 square degrees. Very approximately, one quarter of the 3XMM area overlaps with the SDSS area. This implies that the 3XMM pointings within the SDSS survey area cover, very crudely, around 2\% of that area. The fraction, however, varies strongly with the flux limit, because different observations have different depths and because the sensitivity varies with the location of the source within an XMM pointing. 

In order to estimate the fraction as a function of flux limit (which can then be converted to a function of luminosity and redshift), the following procedure was adopted. The sky density of X-ray sources allows an estimation of the average sky area surrounding each X-ray source according to its flux. By combining this information with the number of 3XMM sources within the SDSS survey area, it is possible to estimate the fraction of the SDSS survey area covered by the XMM pointings, again as a function of flux. After binning the LINERs by flux, their number can be divided by the relevant fraction to permit a like-for-like comparison with the number of SDSS galaxies. The methodology in detail is described below.

The space density of X-ray sources, split by flux, was taken from the parameterisation derived by \citet{geor08}. They found that the number of X-ray sources above a flux level, $\mathrm{f_{x}}$, per square degree was given by the expression:
\begin{equation}
N = \begin{cases}
  \mathrm{-K' \frac{f_{ref}}{1+\upbeta_{2}}\left(\frac{f_x}{f_{ref}}\right)^{1+\upbeta_2}}, & \text{if $\rm{f_{x} < f_{b}}$},\\
  \mathrm{K\frac{f_{ref}}{1+\upbeta_1}\left[\left(\frac{f_b}{f_{ref}}\right)^{1+\upbeta_1} - \left(\frac{f_x}{f_{ref}}\right)^{1+\upbeta_1}\right]}\\ \,\,\,\,\,\,\,- \mathrm{K' \frac{f_{ref}}{1+\upbeta_2}\left(\frac{f_x}{f_{ref}}\right)^{1+\upbeta_2}}, & \text{otherwise},
\end{cases}
\end{equation}
where the normalisation constants K and $\mathrm{K'}$ are related by $\mathrm{K'} = \mathrm{K}(\frac{\mathrm{f_{b}}}{\mathrm{f_{ref}}})^{\upbeta_{1} - \upbeta_{2}}$, $\mathrm{f_{b}}$ is the X-ray flux of the break of the double power law, $\mathrm{f_{ref}} = 10^{-14}$ erg s$^{-1}$ cm$^{-2}$ is the normalisation flux and $\upbeta_{1}$, $\upbeta_{2}$ are the power-law indices for fluxes fainter and brighter than the break flux respectively. For the 2.0 - 10.0 X-ray band, Georgakakis et al (2008) found values for $\upbeta_{1}$, $\upbeta_{2}$, $\mathrm{f_{b}}$ and K of -1.56, -2.52, 10$^{0.09}$ (in units of $10^{-14}$ erg s$^{-1}$ cm$^{-2}$) and 3.79 (in units of $10^{16}$ deg$^{-2}$/(erg s$^{-1}$ cm$^{-2}$)) respectively.

The number of sources per square degree was then inverted to obtain an estimate of the average area surrounding each X-ray source based on its flux. 

All the 3XMM sources within 5 arcminutes of an SDSS spectroscopic source were found. There are just over 70,000 of them and this was taken as a close estimate of the number of 3XMM galaxies within the SDSS survey area. These sources were binned into 47 flux bands, spanning the range 10$^{-15.85}$ to 10$^{-11.15}$ erg s$^{-1}$ cm$^{-2}$. For each flux band, the number of sources was multiplied by the average area per source. It was assumed that, given the wide sky coverage, clustering effects and edge effects would cancel out. The resulting estimate of areal coverage as a function of flux limit is plotted in Figure 2 (the blue points).

Targeting of specific sources could affect this estimation method, and this is expected to be most prevalent at the highest fluxes (as also suggested by the upturn in Figure 2 at these fluxes). This can be tested using the assumption that all those sources within 20 arcseconds of the centre of a pointing were targeted. The green triangles in Figure 2 reveal that at fluxes above 10$^{-13.0}$ erg s$^{-1}$ cm$^{-2}$ the fraction of centred sources rises sharply. Datapoints in the flux bands above 10$^{-13.0}$ erg s$^{-1}$ cm$^{-2}$ were, therefore, discarded when finding the best-fit polynomial. A polynomial of order five was fitted to the datapoints, shown as the blue line. The formula of the line is:
\begin{dmath} \label{eq:rlumvsxlum}
\mathrm{log\left(Area\right)} = -0.001003\,\mathrm{F_X}^{5}-0.04914\,\mathrm{F_X}^{4} - 0.7954\,\mathrm{F_X}^{3}-3.542\,\mathrm{F_X}^{2} +22.91\,\mathrm{F_X}+186.1
\end{dmath}
where the areal coverage of the 3XMM -- SDSS overlap is in square arcminutes and $\mathrm{F_X}$ is the log$_{10}$ flux in erg s$^{-1}$ cm$^{-2}$. The log(Area) was assumed to remain constant above $\mathrm{F_{X} = -12.0}$ erg s$^{-1}$ cm$^{-2}$ , as would be expected at these bright fluxes which should be detectable in all observations.

\begin{figure}
 \includegraphics[width=250px]{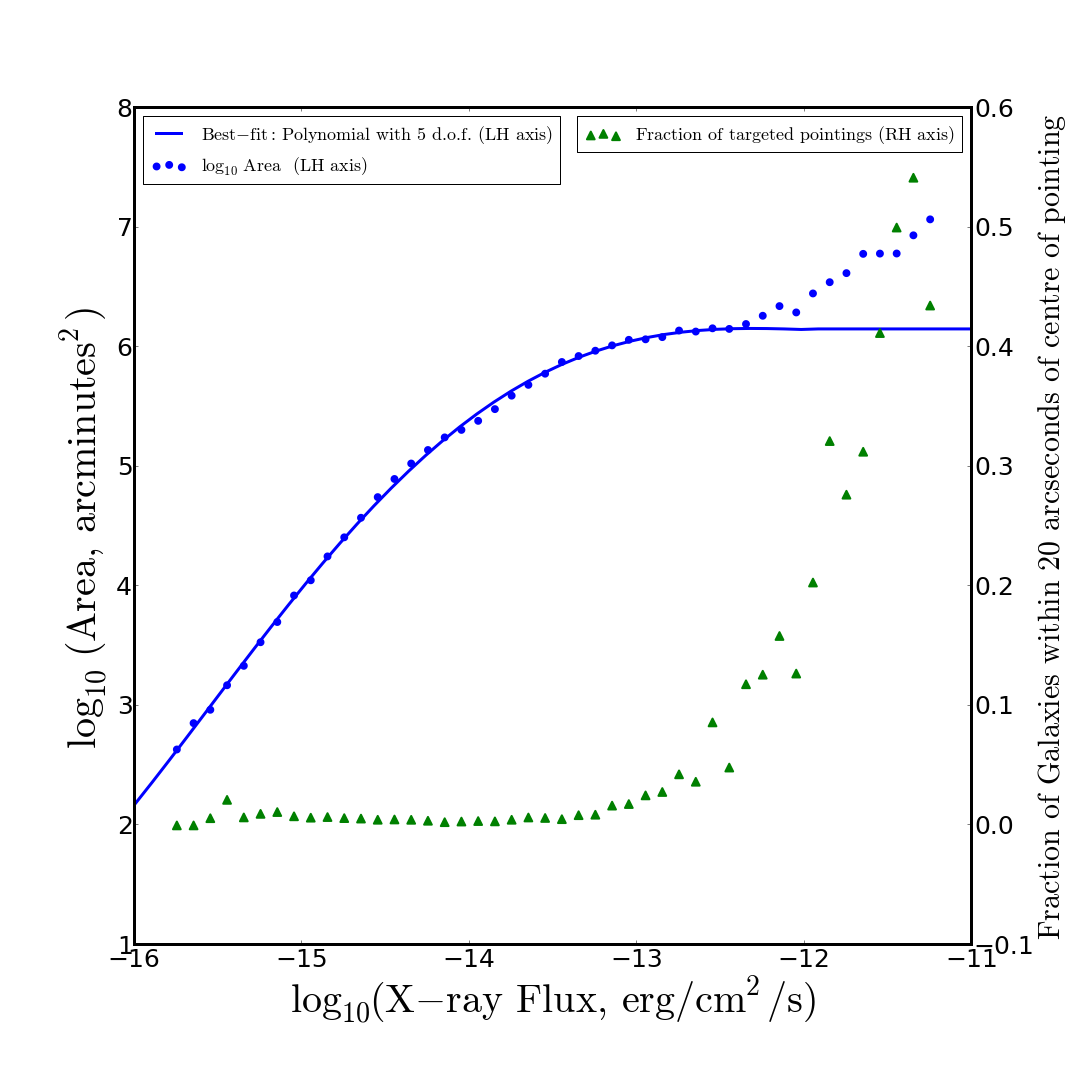}
 \setlength{\abovecaptionskip}{0pt} 
 \caption{The blue dots show the areal coverage of the 3XMM -- SDSS-DR7 overlap region as a function of flux limit. The green triangles, which show the fraction of galaxies within 20 arcseconds the centre of a pointing (right-hand axis), indicate the targeting of those sources with fluxes above $10^{-13}$ erg $\mathrm{cm^{-2} s^{-1}}$. Data points below that flux level are fitted with a polynomial of order 5 (shown by the solid blue line); the polynomial gives the average area surrounding each X-ray source as a function of its flux.}
\end{figure}

The average area that surrounds each X-ray source based on its flux could then be divided by the total area covered by the SDSS survey to find the fraction of areal coverage. So, it is possible to determine, for an SDSS galaxy at a given redshift, what the probability is that it will have been observed in 3XMM pointings with sufficient sensitivity to detect it down to any given luminosity.

\subsection{Deriving the Mass Function}
It was now possible to proceed to derive the mass fraction of LINERs. The fraction of galaxies hosting a LINER at luminosity L was evaluated as:
\begin{dmath} \label{eq:rlumvsxlum}
\mathrm{f_{LINER}(L) = \frac{\sum_z N_{LINER}(z,L)}{\sum_z(fract(z,L) \times N_{SDSS}(z))}}
\end{dmath}
where $\mathrm{N_{LINER}(z,L)}$ is the number of LINERs in the sample at redshift z and X-ray luminosity L, $\mathrm{N_{SDSS}(z)}$ is the number of SDSS galaxies at redshift z and $\mathrm{fract(z,L)}$ is the (luminosity-dependent) areal-coverage fraction that must be applied to ensure a like-for-like comparison between the LINERs and SDSS galaxies. Expression (6) was calculated in narrow bins of luminosity which could then be summed to give the overall fraction of galaxies that host a LINER. In order to perform the calculations, the LINERs and SDSS galaxies were divided into 18 luminosity bins, ranging from 10$^{40.0}$ to 10$^{44.5}$ erg/s in widths of 0.25 dex. In each of the luminosity bins, the LINERs were further subdivided into redshift bins of width z = 0.01. 

Datapoints were considered reliable only if the denominator -- the number of potential hosts -- in a given luminosity bin exceeded 20. This is to minimise any distortion caused by targeting biases. The calculation of the number of potential hosts assumes random pointings. The analysis risks distortion when the number of potential hosts is small and so a single targeted pointing has the potential to affect significantly the result. Hence, datapoints were removed from the analysis if the number of potential hosts in any relevant luminosity bin was below the specified threshold. This had the consequence of removing datapoints at low luminosities. It was checked that the results were insensitive to a change in the threshold to 10 or to 30 potential hosts.

Errors were calculated assuming a Poisson distribution for the number of LINERs. If there were no LINERs in a particular luminosity bin, the upper limit was calculated assuming a single LINER. 

\subsection{Results}
Having developed the procedure to calculate the fraction of galaxies hosting a LINER, the mass-dependent LINER fraction could now be obtained. First, three subsamples of LINERs were constructed -- those with luminosity above 10$^{40.0}$, 10$^{41.0}$ and 10$^{42.0}$  erg/s. These represent, therefore, the lowest X-ray luminosity at which the objects in each subsample could have been detected. Second, the LINERs and SDSS galaxies were further divided into bins according to stellar mass, ranging from 10$^{7.5}$ to 10$^{12.0}$ solar masses. For each of the three luminosity samples, the fraction of galaxies that host a LINER was then calculated for each stellar mass bin in which the number of potential hosts criterion was met.  The results, which are plotted in the upper chart of Figure 3, show that the fraction of galaxies hosting a LINER increases significantly with the stellar mass of a galaxy for all three luminosity limits. 

The slopes of the best-fitting straight lines through the three sets of points are 1.32 $\pm$ 0.05 (L $>$ 40.0 erg/s), 1.89 $\pm$ 0.08 (L $>$ 41.0 erg/s) and 1.70 $\pm$ 0.17 (L $>$ 42.0 erg/s). Although the gradients are not tightly constrained because of the paucity of datapoints and the size of the error bars, it is clear that the fraction of galaxies hosting a LINER is strongly dependent on stellar mass. It can also be seen that at low X-ray luminosities and high stellar masses, the fraction approaches unity -- in other words, essentially all high-mass galaxies host a LINER. Finally, we note that a change of 1 dex in the X-ray luminosity leads to a change in the fraction of approximately 1 dex in the opposite direction.

\begin{figure}
\centering

\subfloat {
  \centering
  \includegraphics[width=1.10\linewidth, height = 160pt]{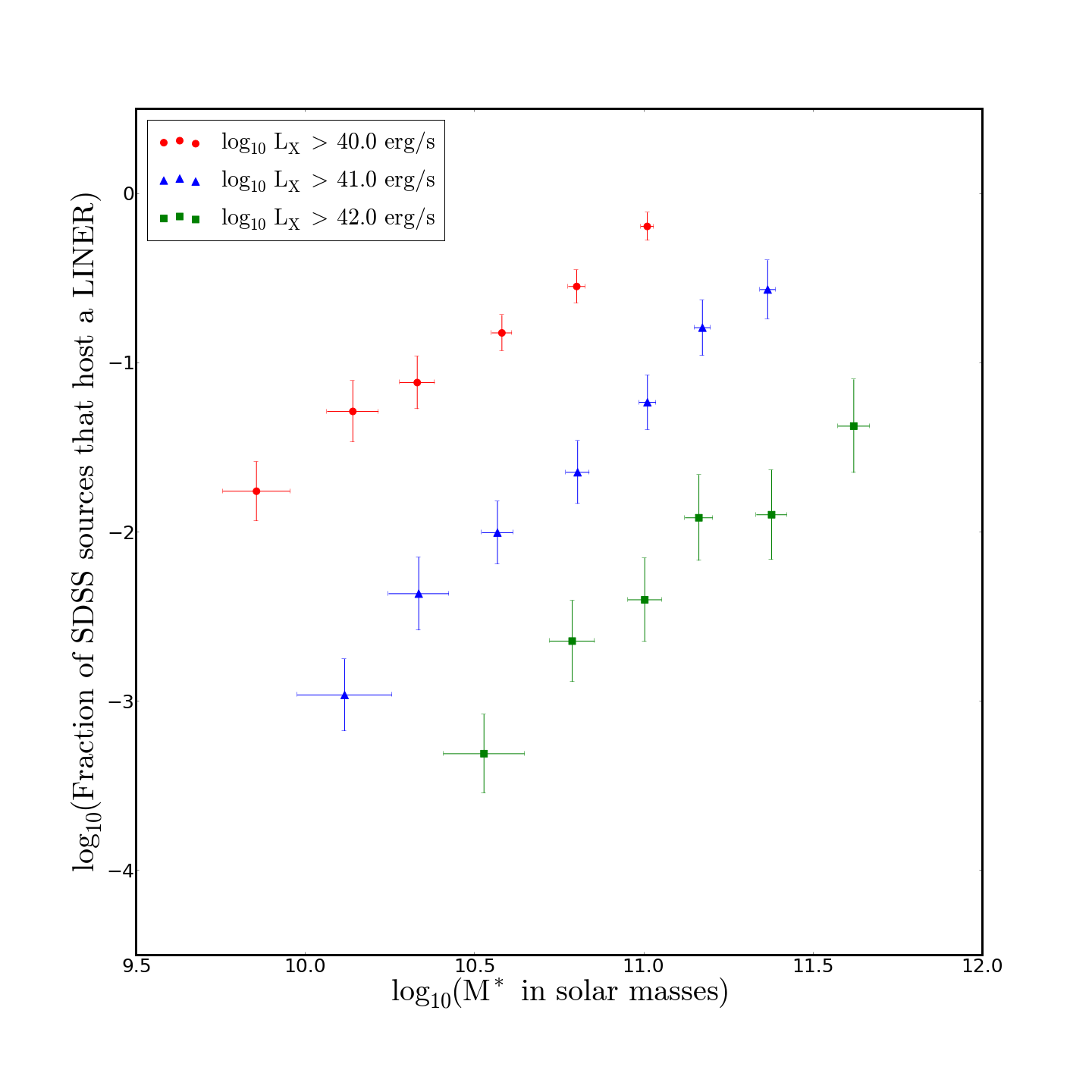}
}

\subfloat {
  \centering
  \includegraphics[width=1.10\linewidth, height = 160pt]{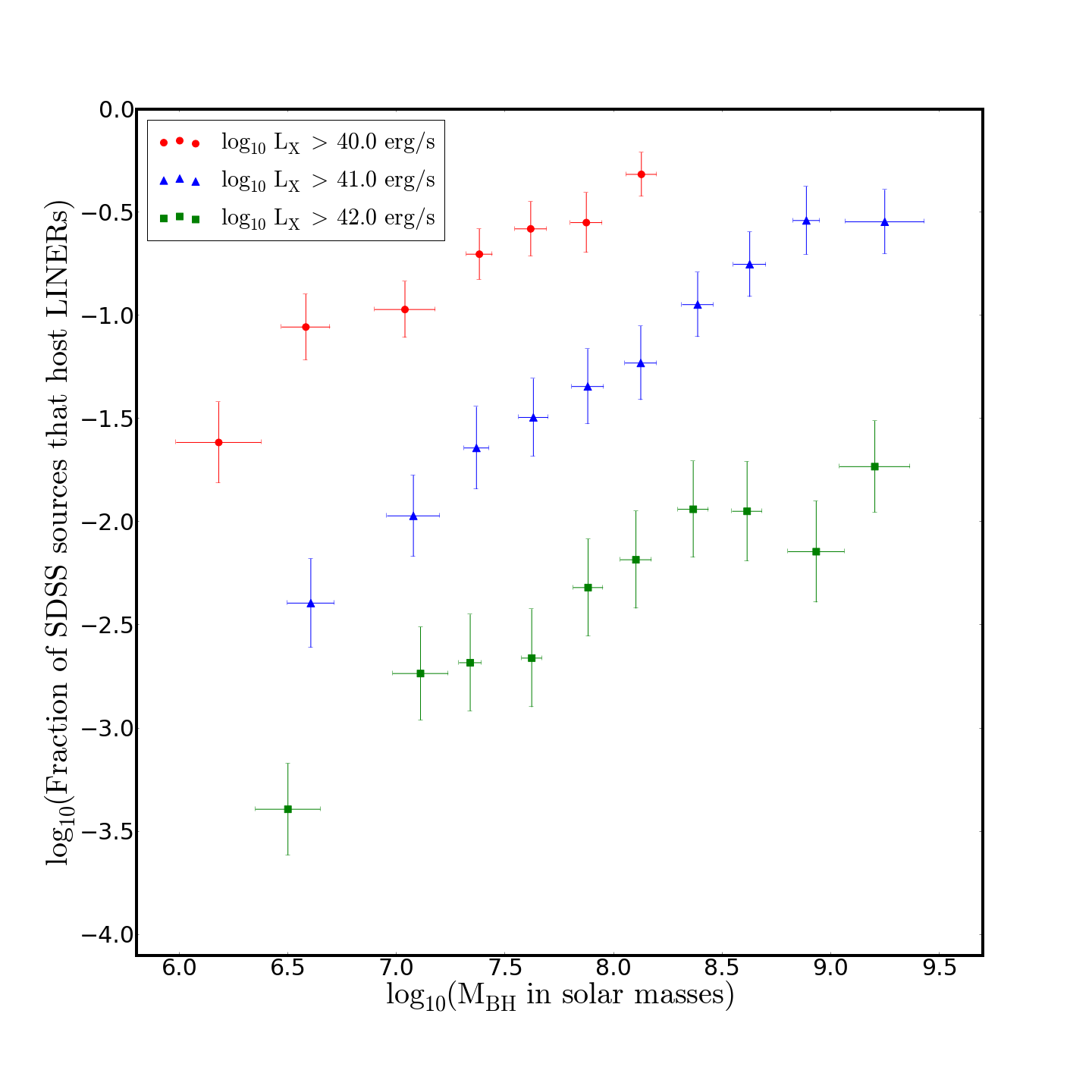}
}

\caption{The fraction of galaxies hosting a LINER according to stellar mass (upper chart) and black hole mass (lower chart). Three subsamples of LINERs are considered - those with X-ray luminosity above 10$^{40.0}$ (red dots), 10$^{41.0}$ (blue triangles) and 10$^{42.0}$ (green squares) erg/s.}
\label{fig:3}
\end{figure}

There have been a number of studies examining the fraction of galaxies displaying X-ray AGN activity. A summary is provided by \citet{bran15} in their Section 5.4. It is of interest to compare the results of these studies with our own, although they are not directly comparable: whereas our research investigated the proportion of galaxies hosting a LINER at redshifts of below 0.30, the other studies have considered the fraction of galaxies hosting any class of AGN within wider redshift ranges. All studies have found, like ours, that the fraction is a strong function of stellar mass. However, the results of these previous studies show flatter slopes and higher normalisations than we find. For example, \citet{aird12} found a slope of $\sim$0.75 that was independent of the X-ray luminosity compared with our slope of 1.6 $\pm$ 0.2. For normalisation, at an X-ray luminosity threshold of 10$^{42.0}$ erg/s and stellar mass of $10^{11.0}$ solar masses, Aird et al found a fraction of $\sim$0.07 compared with our value of $\sim$0.004. Both differences are unsurprising: the fraction of galaxies hosting a LINER is clearly lower than the fraction hosting any AGN, especially at relatively high luminosities where Seyferts dominate; and, given that LINERs tend to reside in older, high-mass galaxies, it is expected that the proportion of galaxies hosting a LINER should have a stronger dependence on stellar mass.

The analysis was repeated, but now using black hole masses rather than stellar masses. Once again, the fraction of galaxies hosting a LINER shows a strong dependence on mass (Figure 3, lower chart), albeit to a lesser extent than for stellar mass. The slopes of the best-fitting straight lines through the three sets of points are 0.57 $\pm$ 0.07 (L $>$ 40.0 erg/s), 0.71 $\pm$ 0.04 (L $>$ 41.0 erg/s) and 0.56 $\pm$ 0.07 (L $>$ 42.0 erg/s). The normalisation remains the same, with an increase of 1 dex in the X-ray luminosity leading to around a 1 dex decrease in the fraction of galaxies hosting a LINER, and vice versa. There is also a faint suggestion that the fractions tend to level off at higher black hole masses, but there are insufficient data to investigate this.

\section{The Fundamental Plane of Black Hole Activity}
It has been postulated that black hole accretion and jet formation is a scale-invariant process; this implies that there will be a relationship between radio luminosity, X-ray luminosity and black hole mass that spans the complete mass range of black holes, from supermassive black holes to stellar X-ray binaries. Such a relationship, defining the Fundamental Plane of black hole activity, was derived empirically first by Merloni et al (2003) and then by a number of researchers (Table 3 below lists a selection). Most have derived the relationship using both AGN and X-ray binaries; a minority have used just AGN, but normally relatively small and heterogeneous samples of AGN.

The scale invariance implied by the Fundamental Plane is valid only when the accretion properties are the same (see Heinz \& Sunyaev 2003 who derive different scalings for different accretion modes). The tightest relationship for X-ray binaries is seen for those genuinely in the ``low-hard'' state, which is for low Eddington fractions. It follows that a valid comparison with ``low-hard'' X-ray binaries requires a sample of jet-mode (LINER) AGN. Our sample of 576 LINERs is large enough to investigate whether the fundamental plane relationship holds within AGN alone.

The distribution of the LINERs' radio flux densities, which is shown in Figure 4, reveals the following points. 196 LINERs (34\% of the total) have a flux density greater than 3$\times$ the rms of the FIRST survey and can, therefore, be considered to be reasonably secure detections. A further 251 LINERs (44\%) have positive flux densities lower than 3$\times$ the rms. 129 of the LINERs (22\%) have a negative flux density. The median and mean flux densities are respectively 0.22 mJy and 6.29 mJy. This implies that there is a lot of information contained in the flux densities of the ``non-detections'' and so it is important to make full use of the data, including low signal / noise detections and indeed negative values. In this case, however, standard stacking techniques are not appropriate because of the wide range of redshifts in the sample. The methodology described below does utilise all of the available information.

\begin{figure}
 \includegraphics[width=250px]{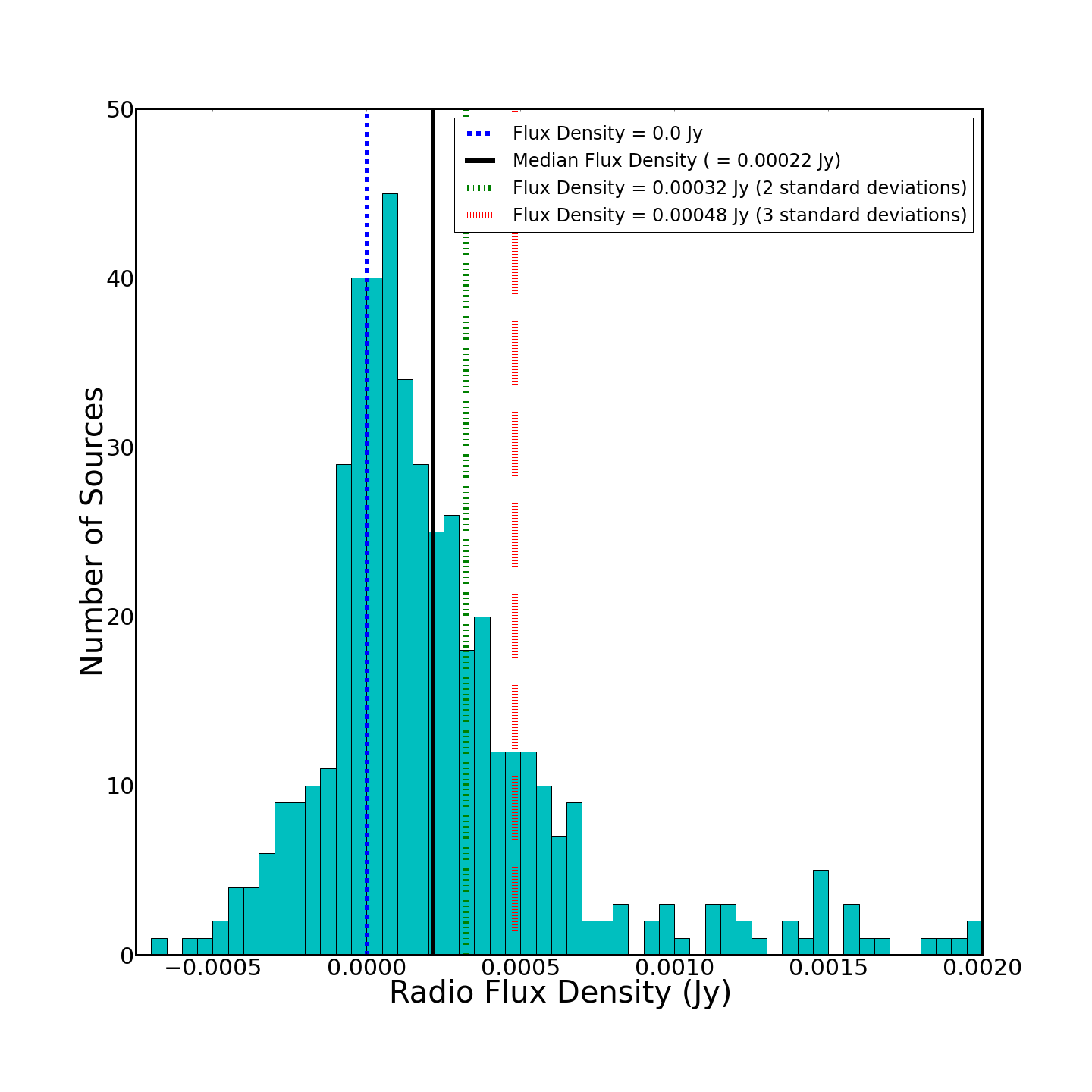}
 \setlength{\abovecaptionskip}{0pt} 
 \caption{The distribution of radio flux densities for the sample of 576 LINERs. Note that 29 LINERs have flux densities greater than 0.002 Jy and are not shown in the diagram. The black dashed line shows the median flux density. The orange and red dashed vertical lines show the flux densities at 2x and 3x the FIRST survey rms respectively.}
\end{figure}

\subsection{The $\rm{L_{X}}$ -- $\rm{L_{R}}$ Relationship}

The relationship between the X-ray and radio luminosities was investigated, first with the black hole mass ignored and then including a dependency on the black hole mass. In deriving the best-fit parameters, two sources of uncertainty in the radio luminosity are considered. The first results from the uncertainty in the measurement of the radio flux densities from the FIRST data, which is assumed to be 0.16 mJy. The second effect to be accounted for is the inherent scatter within the X-ray luminosity / radio luminosity / black hole mass relationship. This scatter arises from both measurement errors due to, for example, source peculiarities, timing differences in the measurements of the X-ray, radio and optical data for variable sources, beaming effects, absorption and the calculation of the black hole masses. It also arises from intrinsic sources of scatter caused by, for example, black hole spin and the fact that the X-ray emission may not be generated solely from the jets. The inherent scatter is unknown and so was modelled as part of the optimisation process.  

Ignoring a possible dependency on black hole mass, the required relationship is expressed in the form:
\begin{dmath}
\rm log\, \left(\frac{L_R}{erg/s}\right)  =  \rm{a}\,\,  \rm{log}\left(\frac{L_X}{10^{42}\,erg/s}\right) + \rm{c}
\end{dmath}
where the denominator under the X-ray luminosity is introduced in order to try and ensure that a and c are largely uncorrelated in the minimisation procedure. The variables to be found are a, c and $\upsigma$ (the inherent scatter in the X-ray luminosity / radio luminosity relationship). The procedure was as follows:
\begin{itemize}
\item For a given value of a, c and $\upsigma$, the predicted radio luminosity of each source was calculated from its X-ray luminosity using the above expression.
\item To allow for the intrinsic scatter in the relation, an addition of N$_1$ times $\upsigma$ was applied to the predicted radio luminosity, where N$_1$ was allowed to vary in small steps between -10.0 and +10.0. So, for each set of values of a, c and $\upsigma$, a large number of predicted radio luminosities were generated.
\item All of these predicted radio luminosities were converted to flux densities. Each one was compared with the actual radio flux density. From this, N$_2$, the number of standard deviations in the offset, was calculated, given the uncertainty in the radio flux density of 0.16 mJy.
\item N$_1$ and N$_2$ were combined in quadrature. The minimum value, $\mathrm{N_{min}}$, for each source was then found for those values of a, c and $\upsigma$. This represents the combined number of standard deviations that the radio flux density is away from the prediction. For bright radio sources the dominant contribution to the offset comes from the intrinsic scatter (i.e. $\mathrm{N_{1}}$), whereas for faint sources it is the uncertainty on the FIRST flux density that dominates.
\item Having determined $\mathrm{N_{min}}$ for each source for a given a, c and $\upsigma$, the best-fit values to those parameters were then found by a maximum likelihood analysis where the log likelihood function is given by:
\begin{dmath}
\mathrm{ln\, L = -0.5\, \sum_i (N_{min, i}^{2} + ln(2\uppi\upsigma^{2}))}
\end{dmath}
where i is the sum over all sources.
\end{itemize}

The values of a, c and $\upsigma$ that maximised the log likelihood function were a = 0.70, c = 38.55 and $\upsigma$ = 0.85, yielding the expression:
\begin{dmath}
\rm{log\, \left(\frac{L_R}{erg/s} \right)} =  \rm{0.70}\,\,  \rm{log}\left(\frac{L_X}{10^{42}erg/s}\right) + \rm{38.55}
\end{dmath}
The errors on the slope and intercept were calculated via the inversion of the Hessian matrix and emerged as $\pm$ 0.09 and $\pm$ 0.12 respectively. The error on the scatter was $\pm$ 0.02. The upper panel of Figure 5 shows the radio luminosity plotted against the X-ray luminosity, with the best-fit shown as the orange line.

\begin{figure}
\centering

\subfloat {
  \centering
  \includegraphics[width=1.10\linewidth, height = 160pt]{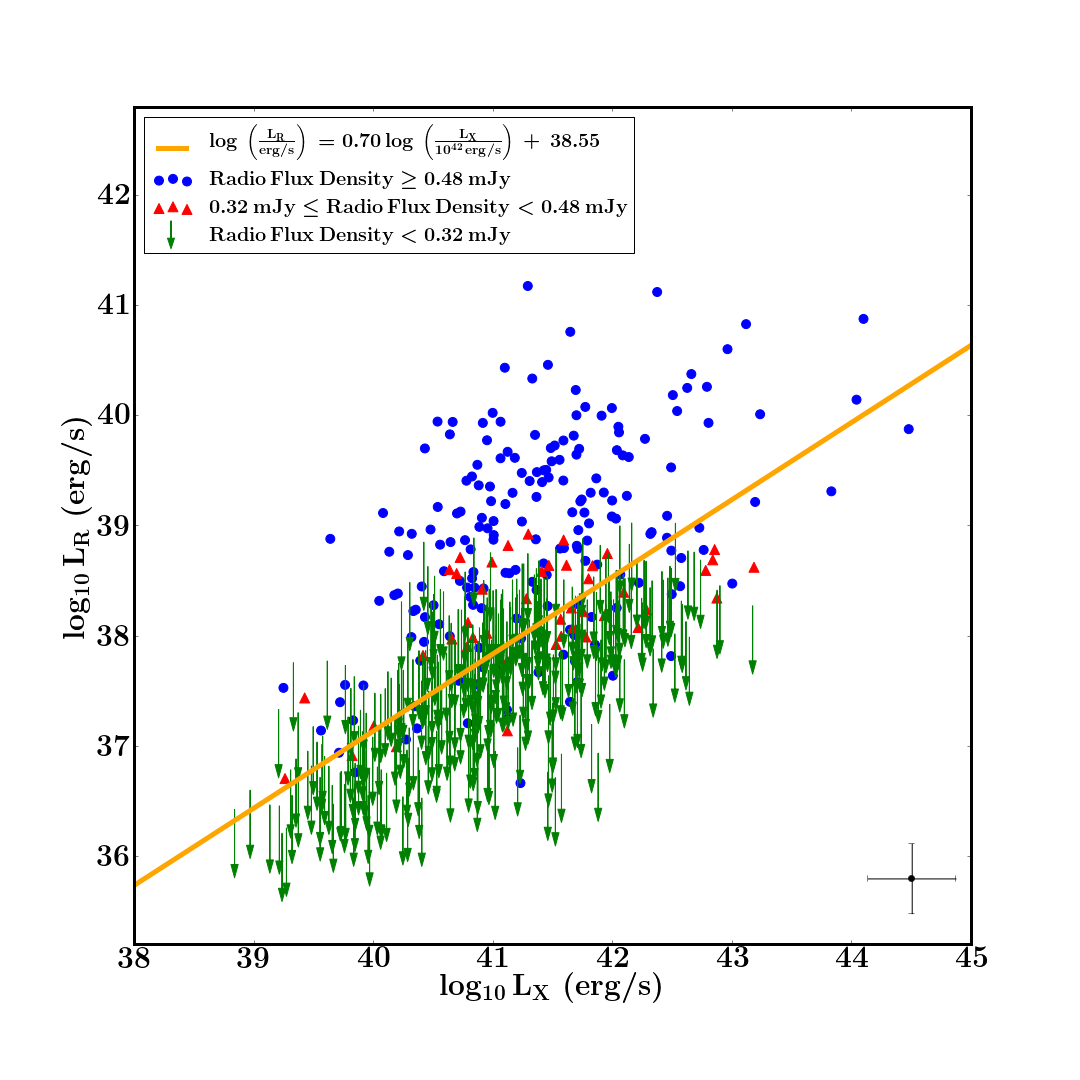}
}

\subfloat {
  \centering
  \includegraphics[width=1.10\linewidth, height = 160pt]{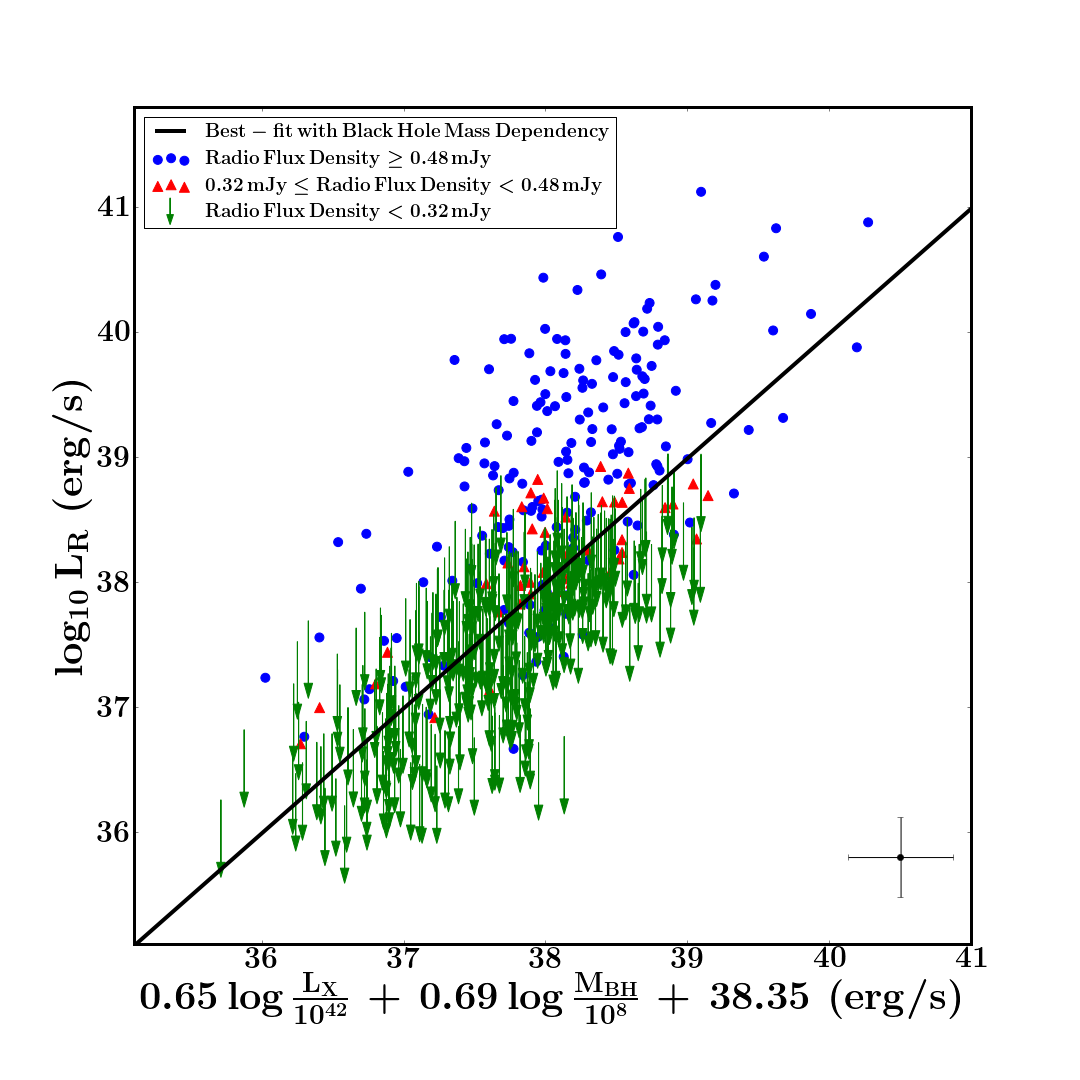}
}

\caption{The Fundamental Plane relationship calculated without black hole mass dependency (upper chart) and with black hole mass dependency (lower chart). In the upper chart, the x-axis shows the log X-ray luminosity and the y-axis shows the log radio luminosity evaluated as $\rm{\upnu\,L_{\upnu}}$. In the lower chart, the x-axis shows the predicted log radio luminosity and the y-axis shows the actual log radio luminosity. In both panels, galaxies with a radio flux density above 0.48 mJy (3$\times$ the FIRST survey rms) have secure radio luminosities and are plotted as blue dots. Galaxies with radio flux densities in the range 0.32 - 0.48 mJy are marginal radio detections and are plotted as red triangles. Galaxies with radio flux densities below 0.32 mJy are plotted as limits, with the base of the arrow corresponding to a radio flux density of 0.32 mJy. Note though, as described in the text, the full radio information was used in the fitting process. Typical error bars are shown in the bottom right of both diagrams.}
\label{fig:5}
\end{figure}

\subsection{The Fundamental Plane}
The exercise was repeated, but now including black hole mass as a dependency. This meant that a fourth variable, the multiple applied to the black hole mass, also needed to be found through the optimisation procedure. The following expression was obtained:
\begin{multline}
\mathrm{log\, \left(\frac{L_R}{erg/s}\right) } =  0.65\,\,  \mathrm{log}\left(\frac{\mathrm{L_X}}{\mathrm{10^{42}erg/s}}\right) \\ + 0.69\,\,\rm{log}\left(\frac{M_{BH}}{10^8\,M_{\odot}}\right)  + 38.35\,\,\,\,\,\,\,\,\,\,\,\,\,\,\,\,
\end{multline}
The variables with their attaching uncertainties were found to be 0.65 $\pm$ 0.07, 0.69 $\pm$ 0.10 and 38.35 $\pm$ 0.10. The value of the scatter, $\upsigma$, emerged as 0.73 $\pm$ 0.03. The best-fit line is plotted as a black solid line in the lower chart in Figure 5.

\begin{center}
\begin{table*}
    \begin{tabular}{| l | c | c | c | c | c | c | c |}
    \hline
Research Paper & \multicolumn{3}{|c|}{Parameters} & \multicolumn{2}{|c|}{Size of Sample} & Description of AGN Sample\\
 & a & b & c & AGN & XRBs & &\\ \hline \hline  
Nisbet \& Best (1) & $0.65^{+0.07}_{-0.07}$ & $0.69^{+0.10}_{-0.10}$ & $38.35^{+0.10}_{-0.10}$ & 576 & 0 & 576 LINERs\\
Nisbet \& Best (2) & $0.70^{+0.09}_{-0.09}$ &  & $38.55^{+0.12}_{-0.12}$ & 576 & 0 & 576 LINERs\\
Nisbet \& Best (3) & $0.65^{+0.05}_{-0.05}$ & $0.96^{+0.10}_{-0.10}$ & $38.42^{+0.07}_{-0.07}$ & 576 & 0 & 576 LINERs\\ \hline
Merloni et al 2003 & $0.60^{+0.11}_{-0.11}$ & $0.78^{+0.11}_{-0.09}$ & 38.42 &  99 & 8 & 14 Quasars, 58 Seyferts, 24 LINERs, 3 Transition Objects\\
K\"{o}rding et al 2006  & $0.63^{+0.4}_{-0.4}$ & $0.75^{+0.3}_{-0.3}$ & 38.47 &  41 & 5 & 41 LINERs\\
K\"{o}rding et al 2006  & $0.57^{+0.07}_{-0.07}$ & $0.78^{+0.18}_{-0.18}$ & 38.01 & 92 & 11 & 14 Quasars, 56 Seyferts, 22 LINERs\\
G\"{u}ltekin et al 2009 & $0.67^{+0.12}_{-0.12}$ & $0.78^{+0.27}_{-0.27}$ & 38.65 & 18 & 0 & 13 Seyferts, 3 Transition Objects, 2 Unclassified Objects\\
G\"{u}ltekin et al 2009 & $0.62^{+0.10}_{-0.10}$ & $0.82^{+0.08}_{-0.08}$ & 38.55 & 18 & 3 & 13 Seyferts, 3 Transition Objects, 2 Unclassified Objects\\
Bonchi et al 2013 & $0.39^{+0.03}_{-0.06}$ & $0.68^{+\rm{n.a.}}_{-\rm{n.a.}}$ & 38.43 & 1268 & 0 & Mixture of both Type 1 and Type 2 AGN\\
Saikia et al 2015  & $0.64^{+0.4}_{-0.4}$ & $0.61^{+0.2}_{-0.2}$ & n.a. & 39 & 4 & 12 Seyferts, 20 LINERs, 7 Transition Objects\\
Saikia et al 2015  & $0.83^{+0.3}_{-0.3}$ & $0.82^{+0.2}_{-0.2}$ & 37.72 & 39 & 0 & 12 Seyferts, 20 LINERs, 7 Transition Objects\\ \hline
    \end{tabular}
    \caption{A selection of Fundamental Plane relationships derived by different researchers. The parameters are defined from: $\mathrm{log\, L_R =  a\,\,  log\left(\frac{L_X}{10^{42}}\right) + b\,\,log\left(\frac{M_{BH}}{10^8}\right)  + c}$. The three models from this research are defined as follows: (1) fitting with black hole mass dependency and using the McConnell \& Ma (2013) relationship between velocity dispersion and black hole mass; (2) fitting without black hole mass dependency; (3) fitting with black hole mass dependency and using the $\rm{M_{BH}\,-\,\upsigma}$ relationship derived by Tremaine et al (2002).} \label{MyTable}
   \end{table*}
   
\end{center}

Including black hole mass as a dependency gave a better fit to the data. The scatter, $\upsigma$, dropped from 0.85 (fitting without black hole mass dependency) to 0.73 (fitting with black hole mass dependency) which indicates that the black hole mass dependency accounts for some of the scatter in the $\rm{L_{X}}$ -- $\rm{L_{R}}$ relationship. A likelihood ratio test was used to investigate whether the improvement is statistically significant once account is taken of the extra degree of freedom. The result was finely balanced. The $\upchi^{2}$ value, computed as -2 $\times$ the natural log of the ratio of the likelihoods, emerged as 2.20. The $\upchi^{2}$ distribution has one degree of freedom, implying that the observed value is at the 86\% confidence interval. This suggests that there is still a 14\% probability that the fit with no black hole mass dependency is the better one. However, as discussed below, the black hole mass dependency is required in order to allow the fundamental plane to fit black holes in a much lower mass range than encountered in LINERs alone (that is, in galactic X-ray binaries).

Table 3 sets out different derivations of the Fundamental Plane relationship. They have all been manipulated into the same format:

\begin{dmath}
\rm{log\, L_R}  =  \rm{a}\,\,  \rm{log}\left(\frac{L_X}{10^{42}}\right) + \rm{b}\,\,\rm{log}\left(\frac{M_{BH}}{10^8}\right)  + \rm{c}
\end{dmath}
and the values obtained for the three parameters are shown in columns 2, 3 and 4 of Table 3. As such, the parameter values derived from different studies should be suitable for comparison since they are largely uncorrelated. There are, however, differences in the definitions of radio luminosity: \citet{merl03}, \citet{kord06} and \citet{gult09} calculated the radio luminosity from an observing frequency of 5 GHz (using $\upnu L_{\upnu}$) and \citet{saik} used an observing frequency of 15 GHz. In this research and that of \citet{bonc13} the radio luminosity was obtained from an observing frequency of 1.4 GHz. We have brought all results in Table 3 into line with an observing frequency of 1.4 GHz. An analysis conducted by \citet{gasp11} on low-luminosity AGN similar to those studied here concluded that $\mathrm{<log\,P_{5 GHz}> - <log\,P_{1.4 GHz}> = 0.35}$ and so a reduction of 0.35 in parameter c was applied to bring the results from Merloni et al, K\"{o}rding et al and G\"{u}ltekin et al onto an equivalent footing to our own. This adjustment corresponds to a mean spectral slope of $\upalpha$ = 0.4, which is reasonable for these small-scale (core-dominated) sources. On the assumption that this spectral index remains relevant up to a frequency of 15 GHz, we have similarly reduced parameter c found by Saikia et al by 0.62. There were also small differences in the methods used by the different researchers to calculate the black hole masses; no attempt has been made to adjust for this. 

\begin{figure}
\centering

\subfloat {
  \centering
  \includegraphics[width=1.10\linewidth, height = 155pt]{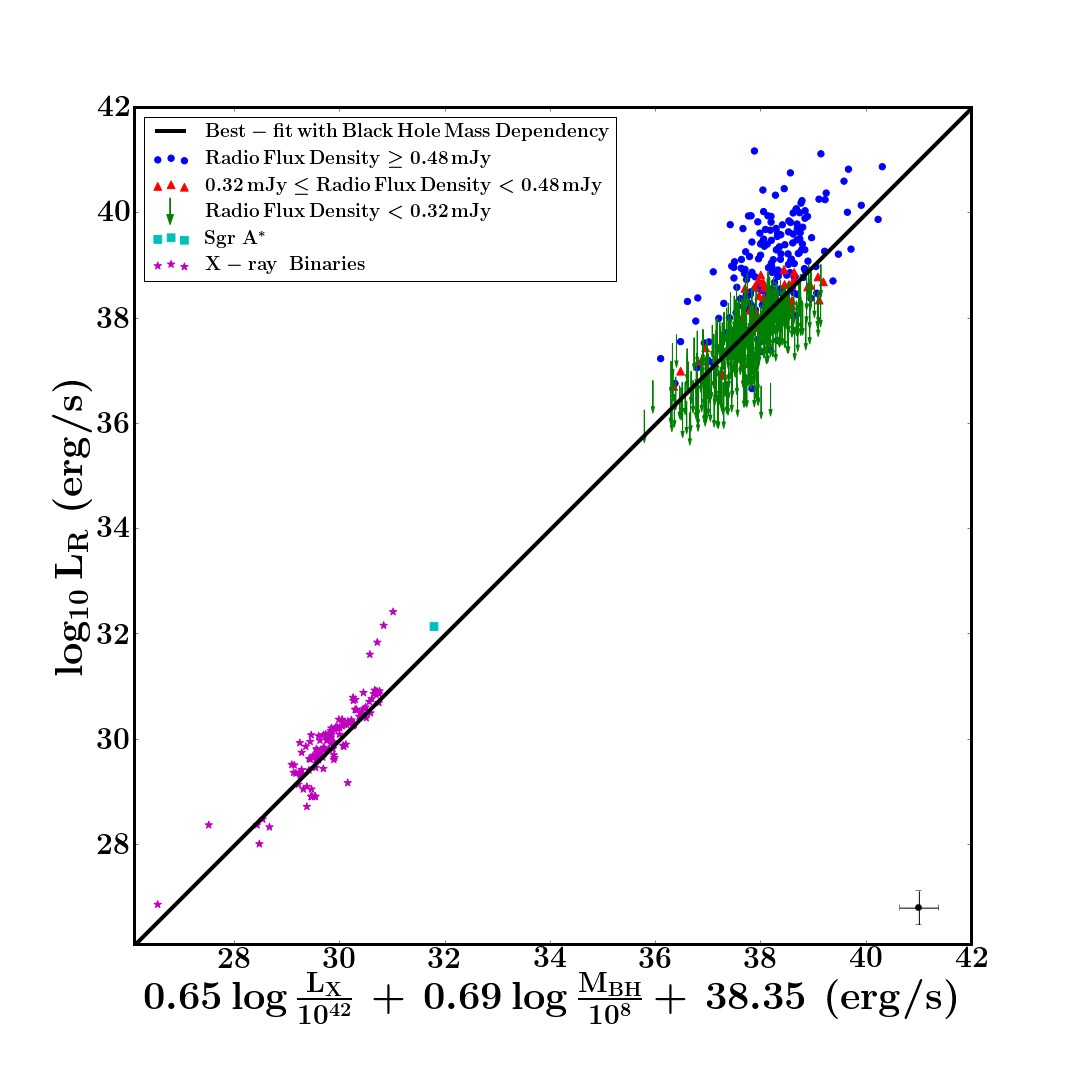}
  }

\subfloat {
  \centering
  \includegraphics[width=1.10\linewidth, height = 155pt]{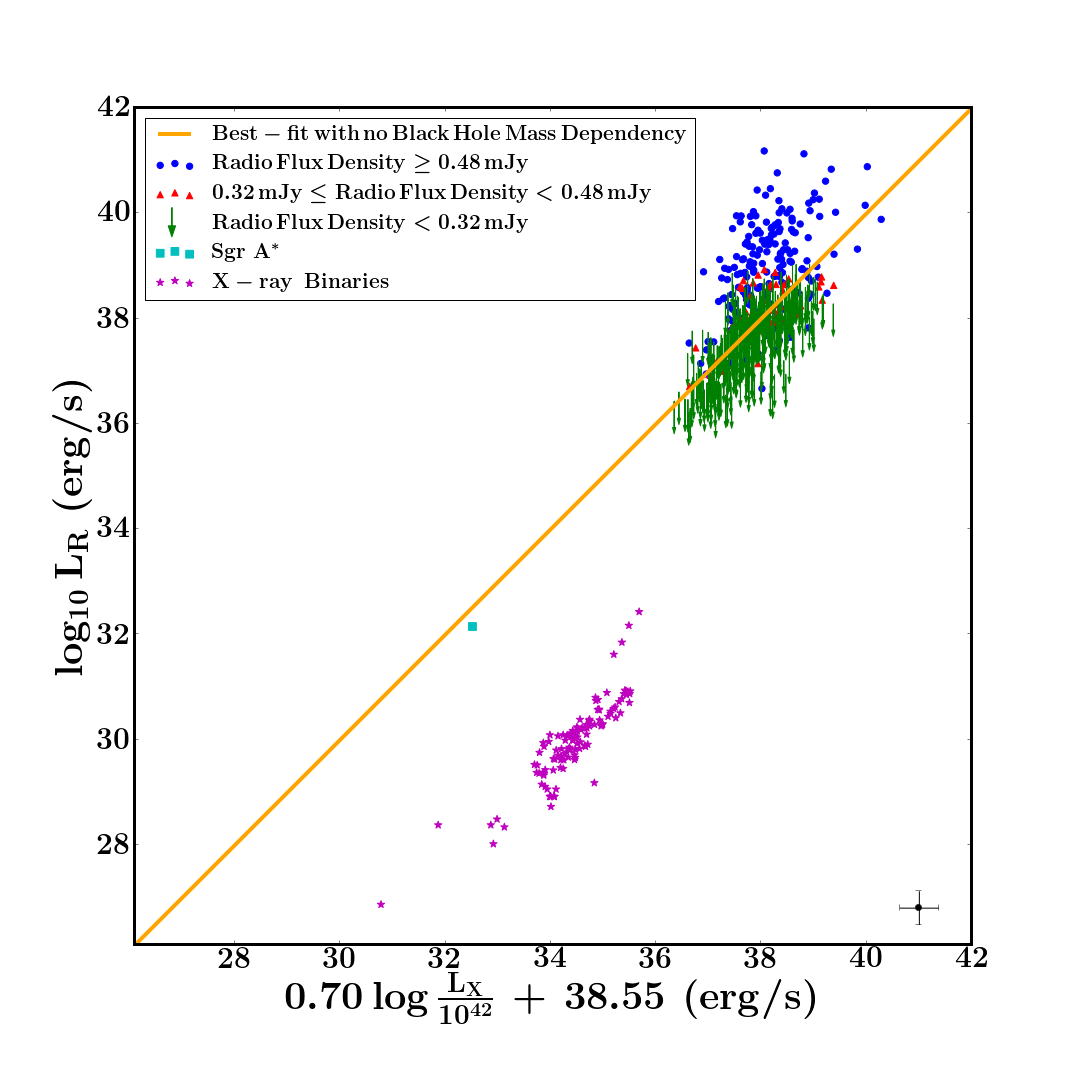}
}
 
\subfloat {
  \centering
  \includegraphics[width=1.10\linewidth, height = 155pt]{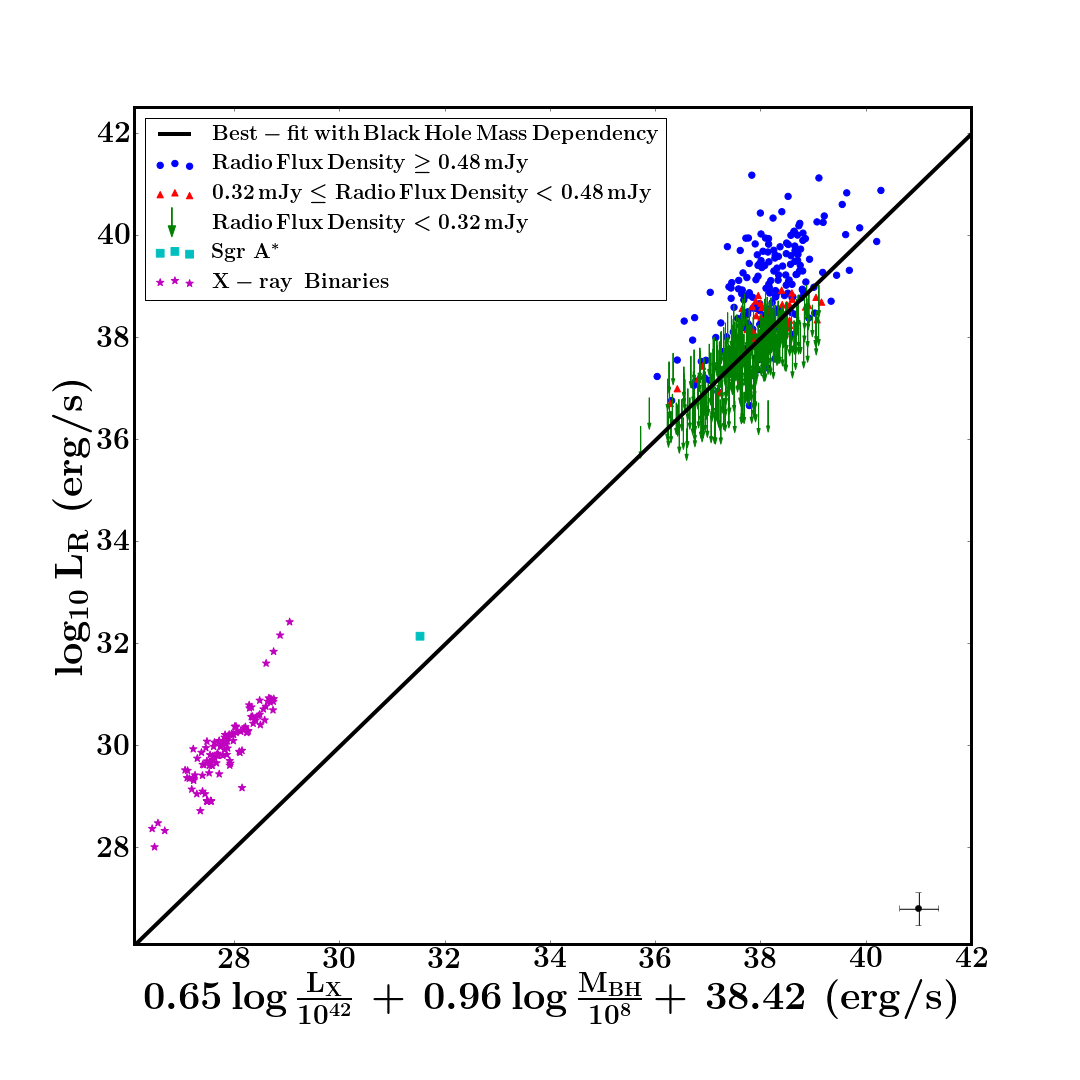}
}

\caption{The observed radio luminosity at 1.4 GHz is plotted against the radio luminosity predicted by the Fundamental Plane relationship. In the upper chart, the relationship includes black hole mass dependency. In the middle chart, the relationship has no black hole mass dependency. In the lower chart, the relationship includes black hole mass dependency, but in this case the black hole masses of the LINERs have been calculated using the black hole mass -- velocity dispersion relationship derived by Tremaine et al (2002). Typical error bars for the LINERs are shown in the bottom right of the three diagrams.}
\label{fig:6}
\end{figure}

K\"{o}rding et al (2006) performed the analysis with a number of different combinations of AGN-type and X-ray binaries; the results of only two are shown here -- a sample that contains only LINERs for the AGN population (and so is consistent with our own research) and a sample that closely replicates that used by Merloni et al (2003). In five cases shown in Table 3, the researchers have derived their relationship using a database consisting of both AGN and also a small number of galactic X-ray binaries (but using several different observations for each of these binaries, which change with time to provide pseudo independent measurements). In the remaining cases, including the Fundamental Plane relationships derived in this paper, only AGN were used in the analysis. 

The parameters derived in this research agree with those in the literature, with our values typically within one standard deviation of those derived elsewhere (Table 3).

A useful test of our relationships is to extrapolate them to lower black hole masses and investigate whether there is agreement with a sample of X-ray binaries. The sample of X-ray binaries was constructed using data from Merloni et al (2003) and Saikia et al (2015). Following the arguments put forward by K\"{o}rding et al (2006), two of the X-ray binaries in the Merloni et al sample were excluded: Cyg X-1, because it changes its state frequently \citep{gall03}, and GRS 1915 + 105, because it seemingly remains for most of its time in a very high state \citep{reig03}. In addition, those datapoints that were limits were excluded. The Saikia sample was included without adjustment. That produced an overall sample of 130 observations from seven X-ray binaries. For comparison, Sgr A* is also included on the plots.

As expected, the Fundamental Plane relationship does need to include the black hole mass dependency in order to fit the X-ray binaries. In Figure 6 the observed radio luminosity is plotted against the predicted radio luminosity, calculated using the derived relationships. The upper chart illustrates that when black hole mass is included as a dependency  the Fundamental Plane does indeed straddle the datapoints of the X-ray binaries for parameters derived only from a fit of the LINER AGN distribution. That is clearly not the case when black hole mass is excluded (middle chart). 

As discussed earlier, the black hole masses have been estimated using the \citet{mcco13} black hole mass -- velocity dispersion relationship. It is of interest to repeat the whole analysis, but now using the relationship derived by Tremaine et al (2002) to calculate the black hole masses. This generates a Fundamental Plane defined by the parameters $a = 0.65^{+0.05}_{-0.05}$, $b = 0.96^{+0.10}_{-0.10}$ and $c = 38.42^{+0.07}_{-0.07}$. The increase in parameter b from 0.69 (using a power of 5.64 in the black hole mass -- velocity dispersion relation) to 0.96 (using a power of 4.02) is consistent with the fact that, given $\rm{M_{BH}\,\propto\,\upsigma^{x}}$, then we expect $\rm{b \propto \,\left(\frac{1}{x}\right)}$. The plane now undershoots the X-ray binaries (Figure 6, lower chart). The analysis provides support for the steeper black hole mass -- velocity dispersion relationship derived by McConnell \& Ma.

\section{Comparison with the Radio-selected LINER fraction}
\citet{best05} demonstrated that the fraction of galaxies hosting radio-loud AGN is a strong function of both stellar and black hole mass. Having derived the Fundamental Plane relationship between X-ray luminosity, radio luminosity and black hole mass, it is now possible to convert from an X-ray selected sample of LINERs to a pseudo-radio-selected sample, so allowing a comparison between the results obtained in Section 4 and those of Best et al.

Again, three subsamples were examined, consisting of LINERs with X-ray luminosities corresponding (according to the Fundamental Plane relationship derived above) to radio luminosities above $10^{38.0}$, $10^{38.5}$ and $10^{39.0}$  erg/s. In practice, this means evaluating down to a different $\rm{L_{X}}$ limit at each black hole mass. The fraction of galaxies hosting each of these subsamples, as a function of black hole mass, was found. The results are shown in Figure 7. The comparison is made against a radio-selected sample of AGN, also from SDSS-DR7, drawn from the analysis of \cite{best12}. For this analysis, their radio sample was restricted to only jet-mode AGN, re-evaluated for only core radio flux and selected above $10^{39.0}$  erg/s. Due to radio flux density limits, the sample had to be restricted to z $\rm{<}$ 0.1, which gives a slightly different redshift coverage to our X-ray selected sample.

A key result is that the slopes found in the two analyses are in reasonable agreement. For the three subsamples in this analysis, the best-fitting straight lines through the datapoints had slopes of 1.9 $\pm$ 0.2  ($\rm{L_{R}\, >\, 10^{38.0}\,\, erg / s}$), 1.7 $\pm$ 0.1 ($\rm{L_{R}\, >\, 10^{38.5}\,\, erg / s}$) and 1.1 red{$\pm$ 0.3} ($\rm{L_{R}\, >\, 10^{39.0}\,\, erg / s}$). We note that these subsamples contain only between four and six datapoints each and so the gradients are not tightly constrained. The best-fitting slope for the jet-mode AGN radio-selected sample is 1.3. There is, however, an indication that for this sample the best-fit turns over at around $10^{9.0}$ solar masses and if a straight line is fitted just through the points below that level then the slope emerges as 1.6. Overall, therefore, the analysis tends to support the conclusions of Best et al (2005), and also those of \citet{jans12}, that the fraction of galaxies hosting a jet-mode AGN scales approximately as $\mathrm{M^{1.6}_{BH}}$. This provides reassurance that the results in Section 4 -- the fraction of galaxies hosting a LINER as a function of black hole mass -- are reasonable, as are the parameter values a = 0.65 and b = 0.69 found for the Fundamental Plane relationship. 

\begin{figure}
 \includegraphics[width=250px]{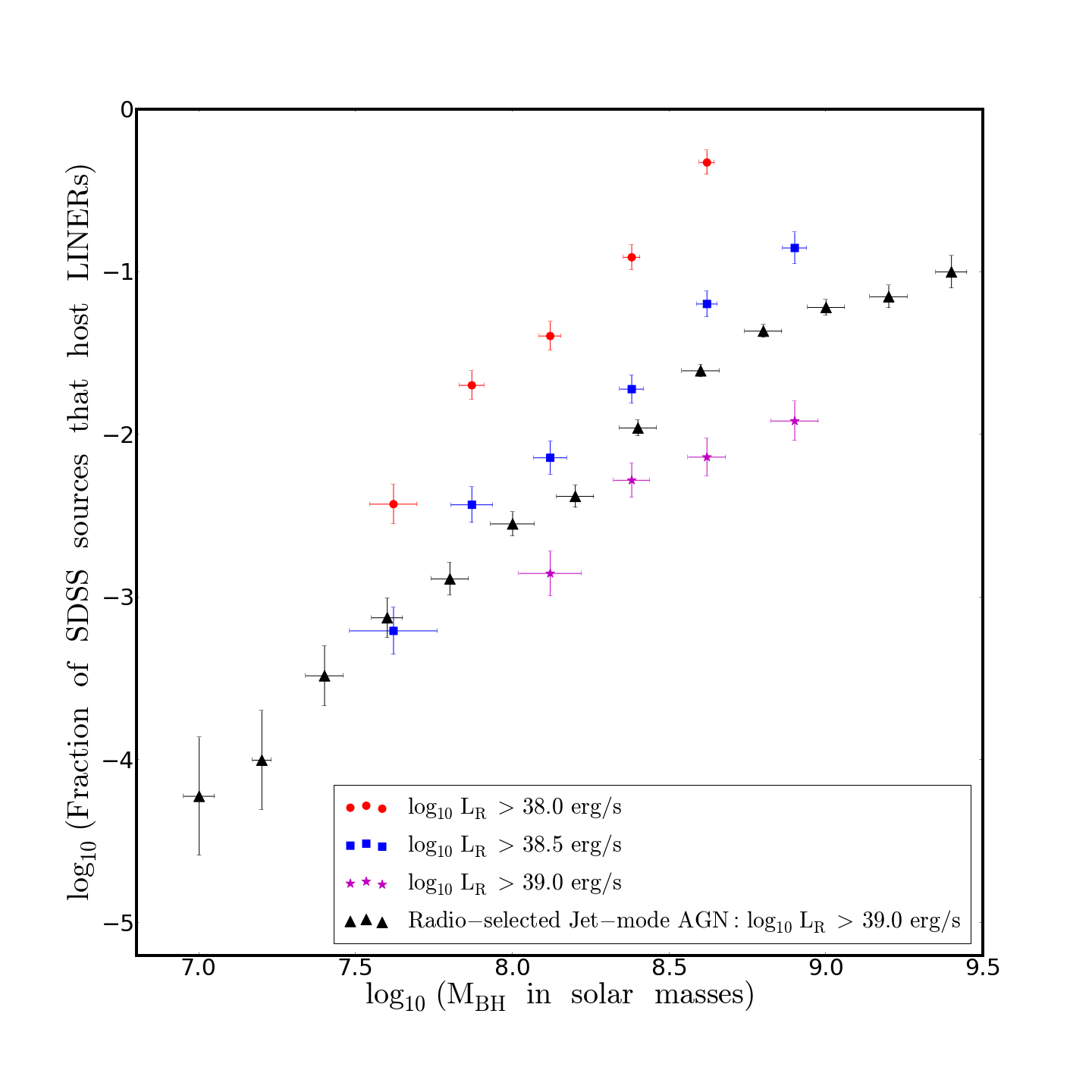}
 \caption{A comparison of the fraction of galaxies hosting LINERs calculated using a radio-selected sample of LINERs (black triangles) with a sample of X-ray-selected LINERs that have been converted into a pseudo-radio-selected sample using the Fundamental Plane relationship (red, blue and magenta symbols). The radio sample consists only of jet-mode AGN and uses core radio luminosities. All samples use the McConnell \& Ma (2013) relationship to estimate the black hole mass.}
\end{figure}

Figure 7 also reveals that there is a small but non-negligible disagreement in the normalisation: results for the radio-selected subsample with radio luminosity above $10^{39.0}$ erg/s are between the results for the subsamples of X-ray selected LINERs with converted radio luminosity above $10^{38.5}$ erg/s and those with radio luminosity above $10^{39.0}$ erg/s, but are slightly closer to the former. There is an offset of approximately 0.3 dex between the two sets of results, suggesting that there may be a discrepancy in the derived value of the intercept in the Fundamental Plane relationship. Even allowing for the fact that small uncertainties in the Fundamental Plane parameters magnify up into changes in the predicted radio luminosity, a discrepancy of 0.3 dex does appear large in the context of the calculated error of 0.10 in the intercept.  

There are several factors that might contribute to the offset. The different redshift regimes (an upper limit of z = 0.30 in this research and upper limit of z = 0.10 for the radio-selected sample) might be a factor. It seems unlikely, though, that this would have a major effect, especially since we found no significant difference in our results when restricting our X-ray-selected sample to redshifts of below 0.20. 

The possible inclusion of misclassified Seyferts in the sample and the possibility of incorrect X-ray matches with the SDSS galaxies could both affect the normalisation. To test the former, those LINERs with a ratio of log([O\Rmnum{3}]/H$\upbeta$) $>$ 3.0 -- that is, those closest to Seyferts in the diagnostic diagrams -- were removed from our sample. There were eight of them. This made no difference to the results. Nor is the latter point likely to be significant. Based on a potential contamination of 24 mis-matched sources (see Section 3.3), the 24 sources with upper limits of radio luminosity most outlying from the $\rm{L_{X}}$ -- $\rm{L_{R}}$ relationship were removed and the analysis repeated. The change in the intercept value was less than the 0.1 dex uncertainty.

The explanation for the disagreement in the normalisations is more likely to lie in one or both of the following factors. First, scatter in the Fundamental Plane relation means that the X-ray cuts and radio cuts are not identical: a vertical (X-ray) cut on Figure 5 generates a subsample of LINERs that mostly lie below the best-fit line, whereas a horizontal (radio) cut will generate a different subsample of LINERs that will be biased to be above the best-fit line. This introduces a discrepancy between the two approaches which could lead to a difference in the results if the scatter is asymmetric. Indeed, for the $L > 10^{39.0}$ erg/s cut studied, this samples the very upper end of the distribution (see the upper panel in Figure 5) where such an asymmetric scatter is seen.

Second, there could be an error in the Fundamental Plane parameters. As discussed above, this would need to be at the $\sim3\upsigma$ level given the formally-calculated uncertainties. As a further investigation of this, the $\mathrm{L_X - L_R}$ relationship was re-examined as a function of $\mathrm{M_{BH}}$. It was assumed that at a given $\mathrm{M_{BH}}$ the slope of the relationship and the scatter were fixed at the best-fit values of 0.65 and 0.73 respectively. The LINERs were then divided into bins according to black hole mass and an optimisation procedure was conducted on each of the binned subsamples; the process was as described in Section 5.1, but in this case only one free parameter, the intercept, needed to be found. 

The results are shown in Figure 8, with the values of the intercept found in each mass bin plotted against black hole mass. The best-fit straight line is plotted in black. The slope of this line is 0.68, which is reassuringly consistent with the slope determined for the Fundamental Plane. However, although the large error bars prevent definitive conclusions, there are indications that the straight line does not fit the data well. The values of the intercept remain fairly constant at black hole masses below $10^{8.0}$ solar masses and then rise steeply thereafter. A number of functions were tested as a fit to the data. Two that provided statistically better fits to the data points than did the straight line (using a reduced chi-squared approach as the test) were: an exponential curve; and a horizontal line up to a break point of a black hole mass equal to $10^{7.9}$ solar masses followed by a straight line with slope of 1.02 above the break point. Interestingly, \citet{gult09} also found a steeper Fundamental Plane slope when they examined high $\mathrm{M_{BH}}$ objects, but later (G\"{u}ltekin et al 2014) found a plane consistent with the X-ray binaries when they included AGN with $\rm{M_{BH} \sim 10^{6}}$ -- that concords with the pattern of datapoints in Figure 8. Further investigation into this matter will require a larger sample and is outside the scope of this report, but for our purposes the key point is that the scatter around the straight line seems larger than the formal 0.1 dex error and this might help to explain the issues seen.

\begin{figure}
 \includegraphics[width=250px]{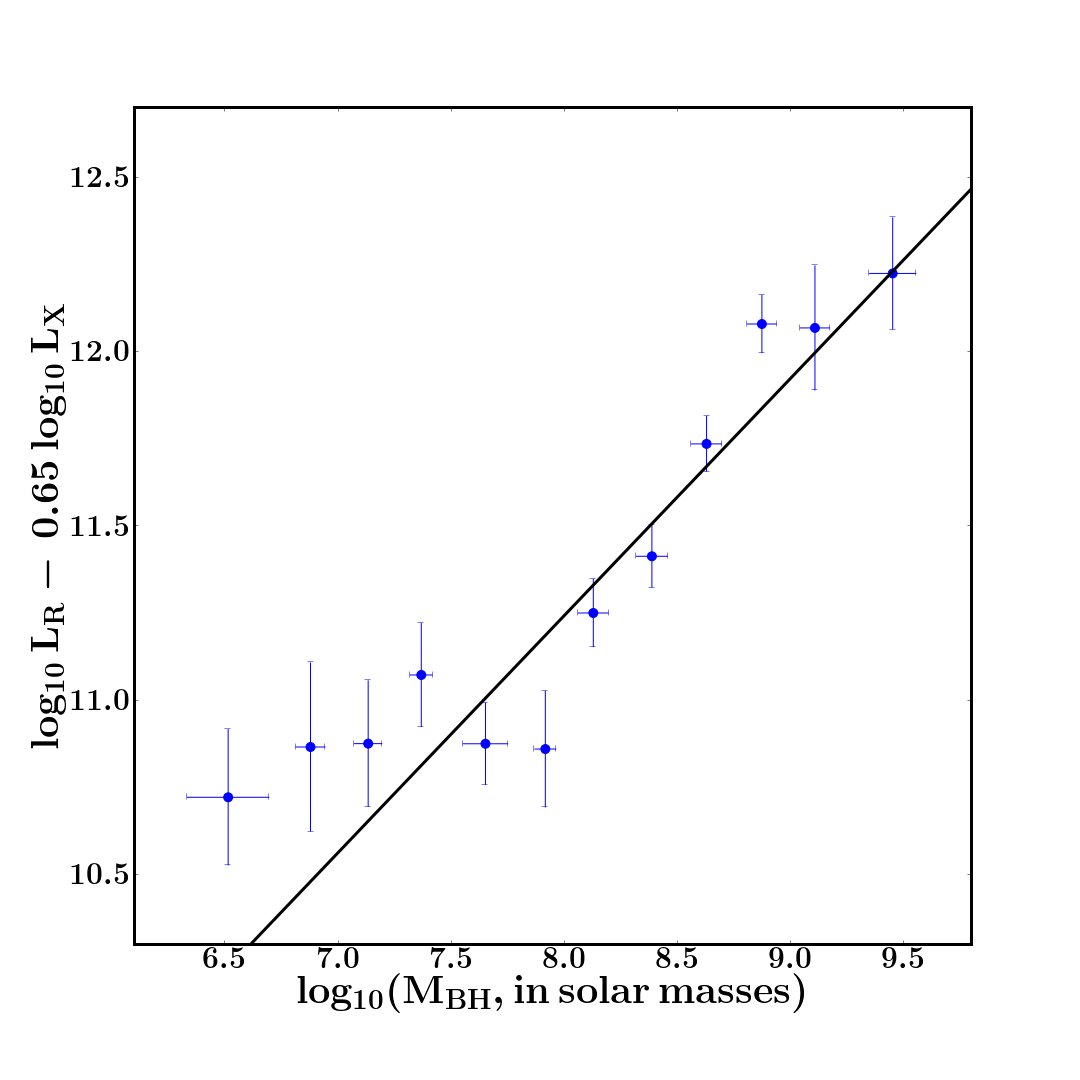}
 \caption{The values of the intercept of the $\mathrm{L_X - L_R}$ relationship have been obtained by dividing the LINERs into bins of black hole mass, keeping parameters a and $\rm{\upsigma}$ constant and then finding the optimal value of parameter c in each bin. The blue points show the results obtained; the best-fit straight line is shown in black.}
\end{figure}

\section{Discussion}
The key points emerging from the results obtained in this research are described in the following subsections.

\subsection{Fundamental Plane of Black Hole Activity}
Most derivations of the Fundamental Plane relationship have included both X-ray binaries and AGN within the analysis. The large difference in both radio and X-ray luminosity between these two subsamples constrains the slope of the plane into a narrow range, implying that the information within the AGN subsamples is not being fully utilised. Similarly, the information contained within the subsamples of X-ray binaries is limited; the number of objects in each subsample is small and multiple observations of the same object are clearly not independent (for example, if black hole spin is a factor). There is also the question of whether the Fundamental Plane relationship is genuine, or whether the fact that the two populations scale together masks variations within these populations. 

Our sample of 576 LINERs has the advantage of being large enough to allow the Fundamental Plane to be derived from the AGN datapoints alone, and then the X-ray binaries can be used as an independent test of the result. Additionally, this sample is restricted to jet-mode AGN, the correct type of AGN to provide consistency with the X-ray binaries. We find a fit that goes through the binaries. The fit is similar to those obtained by other researchers and comfortably within the uncertainties, providing corroboration for their results which have been derived using different approaches to the construction of the sample. We note, however, that our estimate of the slope (parameter a) is towards the top end of the range and our estimate of the dependence on black hole mass (parameter b) is towards the lower end of the range: in any flux-limited sample, these two parameters are expected to be weakly anti-correlated, a fact confirmed by the calculated covariance matrix.

It is of interest that our value of 0.73 dex for the scatter around the relationship, the ``thickness'' of the Fundamental Plane, is in the middle of the range found by other researchers. Such a value suggests that there may indeed be inherent scatter within the relationship, an interesting topic for future research. It has been suggested that the Fundamental Plane could be used as a means of estimating black hole mass from observations of radio and X-ray luminosity (for example, G\"{u}ltekin et al 2009); the size of the scatter appears to limit the practicality of such an approach unless the origin of the scatter can be identified and associated with an observable. We did test a variety of parameters -- [O\Rmnum{3}] line luminosity, the H$\rm{\upalpha}$ line luminosity, the H$\rm{\upbeta}$ line luminosity and the 4000$\mathrm{\mathring{A}}$ breakstrength -- without finding a correlation.

\subsection{XBONGs}
One of the discoveries coming out of recent X-ray surveys is the existence of  X-ray Bright Optically-Normal Galaxies (XBONGs); these are objects that display luminous hard X-ray emission -- typically $\rm{L_{2-10keV}}$ over $10^{42} \rm{erg/s}$ -- but whose optical spectra resemble quiescent normal galaxies with no indication of nuclear activity (for example, Comastri et al 2002 and Trump et al 2009). They, therefore, show the characteristics of an AGN hosted by a “normal” galaxy. A number of explanations have been put forward, including: dilution from the host galaxy light; Compton-thick sources, with X-ray photons having sufficient energy to pass through the torus surrounding the accretion disk, but optical photons unable to do so; and diffuse emission associated with small galaxy groups. One favoured option is that the spectra of XBONGs are due to an advection-dominated accretion flow -- that is, a jet-mode AGN -- in which the X-ray emission, emerging from the base of the jet and the nucleus, has been boosted by a beaming effect along the viewing angle \citep[cf.][]{hart09}.

This research demonstrates that a significant fraction of point-like X-ray sources matched to SDSS galaxies are LINER-like, possessing weak low-ionisation lines. These extend to high X-ray luminosities, with $\mathrm{L_{2-10keV} \sim 10^{42}}$ erg/s. They can, therefore, explain some, or all, of the XBONG population. The research also shows that these sources obey the Fundamental Plane, which implies that there is no evidence of strong beaming. This, in turn, rules out beaming as a general explanation for the XBONG population as a whole. It is possible that the more extreme examples of XBONGs could be boosted by beaming. Indeed, if the origin of the X-ray emission is the jet rather than the advection dominated accretion flow itself, then a proportion of beamed sources would be expected. Further discussion of that topic is beyond the scope of this paper.

We also note that our sample all lie at redshifts of under 0.3, suggesting that dilution is also unlikely to be a general explanation for all XBONGs.

\subsection{AGN Feedback}
The Fundamental Plane relationship implies that the AGN in our sample are experiencing accretion flows with the characteristics of the low-hard state of the X-ray binaries, implying radiatively inefficient accretion flows. This can be investigated by examining the ratio of the mechanical luminosity to the bolometric luminosity. The relationship found for low-luminosity (LINER-like) AGN by \citet{ho09}:
\begin{dmath}
\mathrm{L_{bol} \approx 15.8\,L_{2-10keV}\,\, erg/s}
\end{dmath}
allows an estimate of the bolometric luminosity to be obtained from the X-ray luminosity, while the relationship found by \citet{cava10}: 
\begin{dmath}
\mathrm{L_{mech} \approx 5.8 \times 10^{43}\left(\frac{L_{radio}}{10^{40}}\right)^{0.70} erg/s}
\end{dmath}
allows an estimate of the mechanical energy to be obtained from the radio luminosity. The scatter in both these relationships prohibits too much to be drawn from any simple measurement, but they should provide a reasonable guide to the population. Figure 9 plots the (log) ratio of the mechanical to bolometric luminosity observed for our X-ray selected LINERs. Also shown on the diagram are a series of dashed lines which show the theoretical (log) ratio for a selection of Eddington fractions for sources that follow the Fundamental Plane relationship. These model lines have been calculated by combining equations (10), (12) and (13) with:
\begin{dmath}
\mathrm{f_{Edd} = \frac{L_{mech} + L_{bol}}{L_{Edd}} = \frac{L_{mech} + L_{bol}}{1.26\,\times\,10^{38}\,M_{BH}}}
\end{dmath}
to give:
\begin{dmath}
\mathrm{\left(\frac{L_{mech}}{L_{bol}}\right)^{-0.83} + \left(\frac{L_{mech}}{L_{bol}}\right)^{-1.83} = 1523 \times f_{Edd} \times (M_{BH})^{0.12}}
\end{dmath}
where $\rm{f_{Edd}}$ is the Eddington fraction.
Although there is a lot of scatter and a little over half of all the datapoints are limits (from those LINERs with a radio flux density below 0.32 mJy), the diagram does make several points of interest:
\begin{itemize}
\item For the LINERs, a significant proportion (over half on average for our sample) of the energy is released in the jets. 
\item The data indicate that the fraction of energy released in mechanical form increases with black hole mass -- that is, higher black hole mass objects not only have a higher ``on'' fraction, but are also more efficient at radio-jet feedback. 
\item The negative slope of the model lines shows that, for a fixed Eddington ratio, the ratio of mechanical to bolometric luminosity decreases (gently) with increasing black hole mass. Since the data indicate that the fraction of energy released in mechanical form increases with black hole mass, this implies that black holes with the highest masses tend to have low Eddington ratios.
\item The model lines show that mechanical luminosity becomes progressively more dominant at lower Eddington fractions, as found by, for example, \citet{merl07}.
\item The LINERs mostly lie within a range bounded by Eddington fractions of $10^{-6}$ and $10^{-2}$. This confirms that these sources are in the jet-mode regime, with Eddington fractions of below $10^{-2}$ (cf. Best \& Heckman 2012).
\item Intriguingly, there is a suggestion that the fraction of mechanical to bolometric luminosity rises more sharply above a black hole mass of around $10^{8}$ solar masses. This provides (weak) support for the argument from Figure 8 that some change occurs at that black hole mass threshold, possibly in the accretion flow, in the properties of the black hole or in the properties of the host galaxy. In view of both the large uncertainties in the derivation of the data in Figure 8 and of the scatter within that diagram, we emphasise again that this point is made only in order to indicate an area where further research may be of interest.
\end{itemize}

\begin{figure}
 \includegraphics[width=250px]{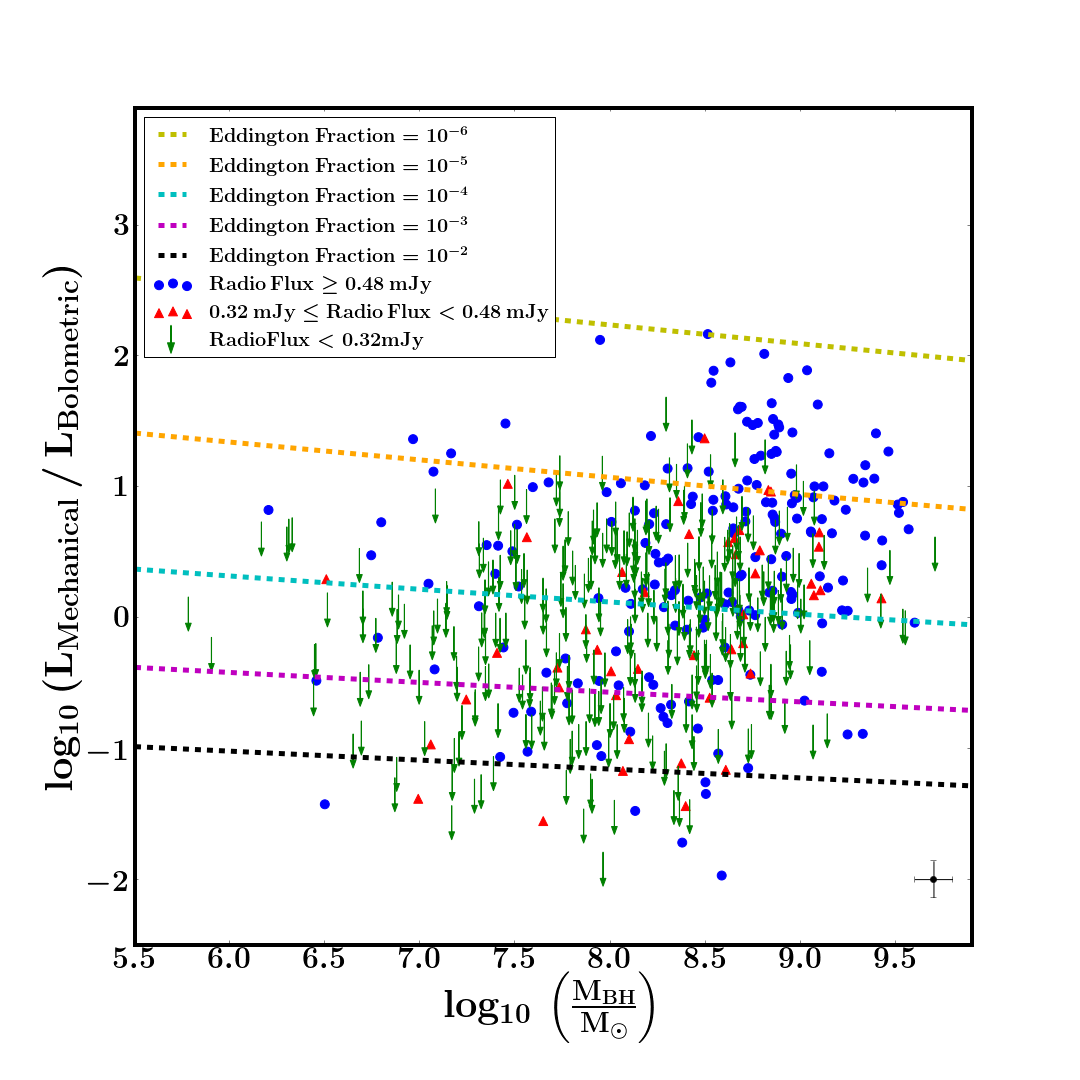}
 \caption{An analysis of the ratio of the mechanical luminosity to the bolometric luminosity, as a function of black hole mass. LINERs with radio flux density above 0.48 mJy (3$\times$ rms) are represented by blue dots. Those with radio flux density between 0.32 mJy and 0.48 mJy (between 2$\times$ rms and 3$\times$ rms) are represented by red triangles. LINERs with radio flux density below 0.32 mJy are represented by green arrows, the base of which is plotted at the luminosity corresponding to 0.32 mJy. The dashed lines show the theoretical ratio as a function of black hole mass for a selection of different Eddington fractions for sources obeying the Fundamental Plane -- the derivation is described in the text. Typical error bars are shown in the bottom right of the diagram.}
\end{figure}

\section{Conclusions}
The main conclusions from this research are:

\begin{itemize}
\item The fraction of galaxies hosting a LINER is a strong function of stellar mass and also of black hole mass. The scaling factors are approximately $\rm{f_{LINER} \propto M^{1.6 \pm 0.2}_{*}}$ and $\rm{f_{LINER} \propto M^{0.6 \pm 0.1}_{BH}}$ respectively. These are shallower than found for radio selection (but are consistent once the Fundamental Plane is considered) and demonstrate that $\rm{\frac{L_{rad}}{L_{X}}}$ scales with $\rm{M_{BH}}$.

\item By studying a sample of jet-mode AGN (radiatively inefficient), we find a Fundamental Plane which matches X-ray binaries in the low-hard state (and so in the same accretion mode as the AGN). The analysis confirms the results from previous research that a scale-invariant Fundamental Plane does exist, and one that spans at least 15 orders of magnitude in X-ray luminosity. The Fundamental Plane was found to be $\rm{log\, \left(\frac{L_R}{erg/s}\right) } =  0.65^{+0.07}_{-0.07}\,\,  \rm{log}\left(\frac{L_X}{10^{42}erg/s}\right) + 0.69^{+0.10}_{-0.10}\,\,\rm{log}\left(\frac{M_{BH}}{10^8\,M_{\odot}}\right) + 38.35^{+0.10}_{-0.10}$.

\item The relationship does, however, have a significant scatter. No simple observable was found that could account for this which argues against the Fundamental Plane being used to estimate accurately black hole mass from observations of X-ray and radio luminosity.

\item For consistency of our Fundamental Plane derivation with the X-ray binary population, our research favours a steep gradient in the relationship between black hole mass and velocity dispersion, as suggested by recent research.

\item It has been argued that the anomalously high X-ray luminosity of XBONGs can be explained by a beaming effect along the line of sight. Our sample of LINERs, which share the same characteristics as XBONGs, display no evidence of strong beaming. The implication is that beaming cannot be a general explanation for all XBONGs.

\item We find that for the LINERs a significant proportion of the energy is released in the jets, and that the proportion rises with increasing black hole mass.

\item Rather more speculatively, it is of interest to note that there are hints of a change in the properties of a black hole or in the accretion flow at a mass of around $10^{8}$ solar masses, as indicated in both Figure 8 and Figure 9.

\end{itemize}

\section*{Acknowledgements}
We thank the anonymous referee for helpful comments and suggestions. We are both grateful for the financial support provided by the STFC. We would like to thank Dr Jos\'{e} Sabater Montes and Dr Bob Mann for their useful comments and, in the former's case, for providing code to assist with the diagnostic process. The research makes use of the SDSS Archive, funding for the creation and distribution of which was provided by the Alfred P. Sloan Foundation, the Participating Institutions, the National Aeronautics and Space Administration, the National Science Foundation, the U.S. Department of Energy, the Japanese Monbukagakusho, and the Max Planck Society. The research makes use of the 3XMM catalogue, prepared by the XMM-Newton Survey Science Centre on behalf of the European Space Agency. The research also uses the FIRST radio survey, carried out using the NRAO VLA: NRAO is operated by Associated Universities Inc., under co-operative agreement with the National Science Foundation.

\nocite{*}
\bibliographystyle{mn2e}

\bibliographystyle{aa}

\section{Appendix}
The properties of a subsample of 20 LINERs are set out in Table 4 on the following page. Data for the full sample of 576 LINERs are available electronically.

\afterpage{
    \clearpage
    \thispagestyle{empty}
\begin{landscape}

\begin{center}
\begin{table}
    \begin{tabular}{| r | r | r | r | r | r | r | c | c | c | r | r | c | r | c |}
    \hline
\multicolumn{1}{c}{Id} & \multicolumn{1}{c}{SDSS} & \multicolumn{1}{c}{SDSS} & \multicolumn{1}{c}{  SDSS} & \multicolumn{1}{c}{RA} & \multicolumn{1}{c}{Dec} & \multicolumn{1}{c}{Redshift} & \multicolumn{1}{c}{Velocity} & \multicolumn{1}{c}{Black Hole} & \multicolumn{1}{c}{Stellar} & \multicolumn{1}{c}{3XMM} & \multicolumn{1}{c}{Xray Flux}  & \multicolumn{1}{c}{Xray} & \multicolumn{1}{c}{Radio Flux} & \multicolumn{1}{c}{Radio}\\

 & \multicolumn{1}{c}{Plate id} & \multicolumn{1}{c}{MJD id} & \multicolumn{1}{c}{Fibre id} & \multicolumn{2}{c}{-------(J2000)-------} & & \multicolumn{1}{c}{Dispersion} & \multicolumn{1}{c}{Mass} & \multicolumn{1}{c}{Mass} & \multicolumn{1}{c}{id} &  & \multicolumn{1}{c}{Luminosity} & \multicolumn{1}{c}{Density} & \multicolumn{1}{c}{Luminosity}\\
 
 & & & & & & & \multicolumn{1}{c}{km/s} & \multicolumn{2}{c}{---$\mathrm{log_{10}\,(in\,\, solar\,\, masses)}$---} & & \multicolumn{1}{c}{$\mathrm{erg/cm^{2}/s}$} & \multicolumn{1}{c}{$\mathrm{erg/s}$} & \multicolumn{1}{c}{Jy} & \multicolumn{1}{c}{$\mathrm{erg/s}$}\\
 
\multicolumn{1}{c}{(1)} & \multicolumn{1}{c}{(2)} & \multicolumn{1}{c}{(3)} & \multicolumn{1}{c}{  (4)} & \multicolumn{1}{c}{(5)} & \multicolumn{1}{c}{(6)} & \multicolumn{1}{c}{(7)} & \multicolumn{1}{c}{(8)} & \multicolumn{1}{c}{(9)} & \multicolumn{1}{c}{(10)} & \multicolumn{1}{c}{(11)} & \multicolumn{1}{c}{(12)} & \multicolumn{1}{c}{(13)} & \multicolumn{1}{c}{(14)} & \multicolumn{1}{c}{(15)}\\ \hline

8 & 928 & 52578 & 162 & 118.1927565 & 25.09057 & 0.0503 & 134.2 & 7.43 & 10.68 & 81323  & 1.864588e-14 & 41.001 & 0.00010601 &    \textit{37.424}\\
9 & 435 &  51882  &   11 & 118.6544565 &  39.17994 & 0.0960 &  191.0 &   8.05 &  11.23 &   82590 & 1.506559e-13 & 42.498 &  0.04881000 &   40.193\\
10 & 544  & 52201  &  127 & 119.6171190 &  37.78657 & 0.0408 & 263.2 & 8.61 &  11.62 & 84098 & 1.668960e-13 &  41.766  & 0.22551000  &  40.086 \\
11 & 2418 &  53794  &  470 & 120.0874095 &  11.31945 & 0.0149 &  112.6 &   7.13 &  10.29 &    77652 & 2.033953e-14 & 39.962 &  -0.00004548 & \textit{36.350}\\
12 & 2267 &  53713  &  209 & 121.4188305 &  15.55045 & 0.0983 &  192.1 &   8.06 &  11.24 &    90736 & 3.744106e-15 & 40.915 &   0.00081525  &  38.438\\
13 & 759 &  52254  &  418 & 122.3806380 &  39.80186 &  0.0639 &  153.9 &   7.67 &  10.95 &    85116 & 3.247637e-14 & 41.458 &   0.00006678  &  \textit{37.639} \\
14 &  440 &  51912  &   73 & 124.2759015 &  48.45555 & 0.1207 & 155.3 &   7.69 &  11.27 &    70901 & 6.941630e-15 & 41.375  & 0.00029003  &  \textit{38.221} \\
15 & 443 &  51873  &  513 & 127.2506490  & 49.98580 & 0.0969 &  204.2 &   8.17 & 10.69 &    36309 & 1.535086e-13 & 42.515 &   0.00014260  &  \textit{38.018} \\
16 & 1930 &  53347 &   259 & 129.4040835 &  24.99980 & 0.0289 &  218.5 &   8.28 &  10.84 &    38438 & 2.521090e-14 & 40.637 &   0.00006993  &  \textit{36.931} \\
17 & 1930 &  53347  &  477 & 129.6000360  & 25.75454 & 0.0182 & 150.8 &   7.64 &  10.39 &   304783 & 1.131406e-13  & 40.880 &  0.09538000  &  38.997 \\
18 & 896 &  52592  &  473 & 131.2109535 &  43.04681 & 0.0569 &  257.8 &   8.57 &  11.36 &   309430 & 1.651034e-14 & 41.059 &   0.00052783  &  37.752 \\
19 & 1210 &  52701  &  108 & 133.2761235 &  33.37857 & 0.0861 &  121.1 &   7.25 &  10.76 &   308323 & 9.738500e-14 & 42.208 &   0.00047685  &  38.083 \\
20 & 2431 &  53818 &  165 & 133.4338230 &  15.13472 & 0.0710 &  169.0 &   7.84 &  10.91 &   308490 & 8.474410e-15 & 40.970 &  -0.00022903 &  \textit{37.734} \\
21 & 471 &  51924  &  125 & 136.6025235  &  0.96556 & 0.0703 &  166.1 &  7.81 &  10.48 &   298764 & 2.608380e-13 & 42.450 &   0.00752107 &   39.097 \\
22 & 553 &  51999  &  523 & 138.3578955 &  52.98131 & 0.0252 &  220.5 &   8.30 &  11.10 &    51714 & 8.160800e-15 & 40.027 &   0.00000774  &  \textit{36.811} \\
23 & 1193 &  52652   &   3 & 138.8849940 &   4.50847 & 0.0981 &  217.5 &   8.28 &  11.10 &   303479 & 1.291340e-14 & 41.451 &   0.00056695 &   38.278 \\
24 & 1936 &  53330  &  621 & 138.9936675 &  29.73059 & 0.1866 &  205.4 &   8.18 &  11.18 &   301504 & 3.176420e-16 & 40.449 &   0.00030246  &  \textit{38.630} \\
25 & 1941 &  53386  &  256 & 141.5714265 &  30.81864 & 0.0756 &  96.9 &   6.87 &  10.71 &   318959 & 5.556300e-15 & 40.845 &   0.00021338  &  \textit{37.792} \\
26 & 486  & 51910   & 426 & 143.8266450  & 61.47545 & 0.1243 &  231.5 &   8.39 &  11.29 &   317810 & 6.276680e-15 & 41.358 &  0.00048021   & 38.424 \\
27 & 570 &  52266   & 267 & 145.1923680   & 3.41278  & 0.0732 &  189.0 &   8.03 &  11.05 &   321767 & 2.830711e-14 & 41.522 &   0.00046989  &  37.929 \\
 \hline
    \end{tabular}
    
    \captionof{table}{A subsample of 20 out of the full sample of 576 LINERs. The column descriptions are as follows:\newline (1) Identifier in this sample of LINERs.
\newline (2) Plate identifier for the SDSS source in the SDSS-DR7 catalogue.
\newline (3) MJD identifier for the SDSS source in the SDSS-DR7 catalogue.
\newline (4) Fibre identifier for the SDSS source in the SDSS-DR7 catalogue.
\newline (5) Right Ascension for the SDSS source.
\newline (6) Declination for the SDSS source.
\newline (7) Redshift derived from the SDSS spectra.
\newline (8) Stellar velocity dispersion of the galaxy bulge, derived from SDSS data, in units of km/s.
\newline (9) $\mathrm{log_{10}\,(M_{BH}\,\, in\,\, solar\,\, masses)}$; black hole mass derived from the velocity dispersion using the McConnell \& Ma (2013) relationship (but see text).
\newline (10) $\mathrm{log_{10}\,(M^{*}\,\, in\,\, solar\,\, masses)}$; stellar mass derived from the SDSS data.
\newline (11) Identifier of X-ray source in the 3XMM-DR4 catalogue.
\newline (12) X-ray flux in the 2.0 - 12.0 keV band, in units of $\mathrm{erg/cm^{2}/s}$.
\newline (13) X-ray luminosity corresponding to the flux in the 2.0 - 10.0 keV band, in units of $\mathrm{erg/s}$.
\newline (14) Radio flux density at an observing frequency of 1.4 GHz, in units of Jy.
\newline (15) Radio luminosity from an observing frequency of 1.4 GHz (using $\upnu L_{\upnu}$), in units of $\mathrm{erg/s}$. If the radio flux density is under 0.32 mJy (2x the FIRST survey rms), the luminosity is shown as a limit corresponding to a flux density of 0.32 mJy and is italicised.}
   \end{table}
   \end{center}
\end{landscape}
\clearpage
}

\label{lastpage}

\end{document}